\RequirePackage{fix-cm}
\documentclass{svjour3}                     
\smartqed  
\usepackage{graphicx}
\usepackage{amsmath,amssymb}
\usepackage{diffcoeff}
\usepackage{arydshln,mathtools}
\usepackage{mathrsfs}
\usepackage{bm}
\usepackage{hyperref}

\newcommand{\secref}[1]{\S\ref{#1}}

\DeclareMathOperator{\Tr}{Tr}

\DeclareMathOperator*{\Grad}{Grad}
\DeclareMathOperator*{\Div}{Div}

\newcommand{\crmat}[1]{\ensuremath{[#1]_{\times}}}

\def\onedot{$\mathsurround0pt\ldotp$}
\def\cddot{
	\mathbin{\vcenter{\baselineskip.67ex
			\hbox{\onedot}\hbox{\onedot}}%
}}

\makeatletter \renewcommand\d[1]{\ensuremath{%
		\;\mathrm{d}#1\@ifnextchar\d{\!}{}}}
\makeatother

\usepackage{etoolbox}
\newcommand*{\affaddr}[1]{#1} 
\newcommand*{\affmark}[1][*]{\textsuperscript{#1}}

\graphicspath{{./Figures/}}

\begin{document}

\title{Port-Hamiltonian flexible multibody dynamics\thanks{This work is  supported by the project ANR-16-CE92-0028, entitled {\em Interconnected Infinite-Dimensional systems for Heterogeneous Media}, INFIDHEM, financed by the French National Research Agency (ANR) and the Deutsche Forschungsgemeinschaft (DFG). Further information is available at {\url{https://websites.isae-supaero.fr/infidhem/the-project}}.}
}


\author{Andrea Brugnoli\affmark[1]\and
	Daniel Alazard\affmark[1] \and 
	Valérie Pommier-Budinger\affmark[1] \and
	Denis Matignon\affmark[1]
}
\authorrunning{
	Andrea Brugnoli  \and
	Daniel Alazard \and 
	Valérie Pommier-Budinger \and
	Denis Matignon
}

\institute{Andrea Brugnoli \at
	\email{andrea.brugnoli@isae.fr} 
	\and
	Daniel Alazard \at
	\email{daniel.alazard@isae.fr} 
	\and
	Val\'erie Pommier-Budinger \at
	\email{valerie.budinger@isae.fr} 
	\and
	Denis Matignon \at
	\email{denis.matignon@isae.fr} \vspace{3mm} \\
	\affaddr{\affmark[1]ISAE-SUPAERO, Universit\'e de Toulouse, France.\\
		10 Avenue Edouard Belin, BP-54032, 31055 Toulouse Cedex 4.}
}


\maketitle

\begin{abstract}
A new formulation for the modular construction of flexible multibody systems is presented. By rearranging the equations for a flexible floating body and introducing the appropriate canonical momenta, the model is recast into a coupled system of ordinary and partial differential equations in port-Hamiltonian (pH) form. This approach relies on a floating frame description and remains valid under the assumption of small deformations. This allows including mechanical models that cannot be easily formulated in terms of differential forms. Once a pH model is established, a finite element based method is then introduced to discretize the dynamics in a structure-preserving manner. Thanks to the features of the pH framework, complex multibody systems are constructed in a modular way. Constraints are imposed at the velocity level, leading to an index 2 quasi-linear differential-algebraic system. Numerical tests are carried out to assess the validity of the proposed approach.

\end{abstract}

\section{Introduction}
\label{intro}
\indent In structural control co-design of flexible multibody systems, it is especially useful to dispose of a modular description, to simplify analysis. In this spirit, the transfer matrix method \cite{Rui2005} and the component mode synthesis \cite{HurtyCMS} are two well known substructuring techniques that allow the construction of complex multibody systems by interconnecting subcomponents together. A reformulation of the Finite Element-Transfer Matrix (FE-TM) method \cite{TAN199047} allows an easy construction of reduced models that are suited for decentralized control design. For the component mode synthesis, the controlled component synthesis (CCS), a framework for the design of decentralized controller of flexible structures, has been proposed in \cite{YoungCMS}. Another modeling paradigm based on the component mode synthesis is the two-input two-output port (TITOP) approach \cite{TITOP}. It conceives the dynamical model of each substructure as a transfer between the accelerations and the external forces at the connection points. This feature allows considering different boundary conditions by inverting specific channels in the transfer matrix. A rigorous validation was provided in \cite{Perez,SANFEDINO2018128}, where the robustness of the methodology in handling various boundary conditions was assessed. \\
\indent The Lagrangian formulation is the most commonly used methodology to retrieve the equations of motion of flexible multibody systems. Nevertheless, the port-Hamiltonian (pH) framework \cite{bookPHs} has been recently extended to describe the dynamics of rigid and flexible links \cite{macchelli_fl,macchelli_flrig}. PH systems are intrinsically modular \cite{CerveraIntFinite}, hence this approach naturally allows constructing complex system by interconnecting together atomic elements. The formulation therein naturally accounts for the non-linearities due to large deformations.  However, this methodology relies on Lie algebra and differential geometry concepts and requires non standard discretization techniques \cite{Golo}. Thus, the overall implementation is not straightforward. \\
\indent Together with the approach used to derive the equations of motion, the incorporation of the elastic motion represents another important point when dealing with flexible multibody systems. Three descriptions are commonly used: the floating frame formulation, the corotational frame formulation and the inertial frame formulation \cite{Ellenbroek2018}. The choice greatly depends on the foreseen application.  The corotational and inertial frame formulations take into account large deformations of the elastic body, hence are well-suited for accurate simulations. Unfortunately, the application of linear model reduction techniques remains impractical \cite{Noor_rev} and the inclusion of active control strategies is often unfeasible due to the computational burden. The floating frame formulation is less accurate but easily integrates many model reduction techniques \cite{NOWAKOWSKI201240}, making it possible to obtain a low-dimensional problem for control design. \\
\indent In this paper, we propose to combine the pH framework with a floating frame description of the dynamics. Starting from the general equation for the rigid flexible dynamics of a floating body, an equivalent port-Hamiltonian system is found by appropriate selection of the canonical momenta. The flexible behavior is based on the linear elasticity assumption making it possible to include models that cannot be easily formulated in terms of differential form  \cite{BRUGNOLI2019940,BRUGNOLI2019961}. The problem is then written as a coupled system of ordinary and partial differential equations (ODEs and PDEs), extending the general definition of finite-dimensional port-Hamiltonian descriptor systems provided in \cite{mehrmann2019structurepreserving}. A suitable structure-preserving discretization method, based on \cite{cardoso2019partitioned}, is then used to obtain a finite-dimensional pH system. The modularity feature of pH systems makes the proposed approach analogous to a substructuring technique \cite{substructuring}: each individual component can be interconnected to the other bodies using standard interconnection of pH systems, as it is done in \cite{macchelli_flrig}. This feature allows the use of modeling platforms like \textsc{Simulink}$^{\tiny{\textregistered}}$ or \textsc{Modelica}$^{\tiny{\textregistered}}$. The constraints are imposed on the velocities, leading to a quasi-linear index 2 differential-algebraic port-Hamiltonian system (pHDAE) \cite{phd_steinbrecher,beattie2018linear}. In the linear case, the algebraic constraints can be eliminated, preserving the overall pH structure, using null space methods \cite{nullspaceFlMult}.  As a floating frame formulation is used, model reduction techniques can be employed to lower the computational complexity of the model \cite{phode_red,phdae_red}. These peculiarities make the proposed formulation interesting for control applications, that can benefit from the properties of pH systems \cite{PHadaptive,ORTEGAsurvey}. \\
\indent The paper is organized in the following manner. In Section \ref{sec:class_model} the classical equations of a flexible floating body, derived by means of the virtual work principle \cite{MB_Daepde,simeon2013computational}, are recalled. Using the properties of the cross product, the equations are recast in a form closer to the pH structure. Section \ref{sec:pH_fd} details the pH formulation of a floating flexible body by introducing the proper canonical momenta. In Section \ref{sec:discr} a finite element based discretization is detailed for the elastodynamics problem. The procedure is easily applied to flexible floating bodies. The particular case of thin planar beams is then detailed, as it will be next employed in the simulation part. Section~\ref{sec:MB_pH} explains how to interconnect models together using classical pH interconnection. Section \ref{sec:valid} is devoted to numerical examples, to assess the validity of the proposed methodology. The test cases are taken from previously published articles \cite{Chebbi2017,Ellenbroek2018}.

\section{Flexible dynamics of a floating body}
\label{sec:class_model}
\begin{figure}[t]
	\centering
	\includegraphics[width=0.6\textwidth]{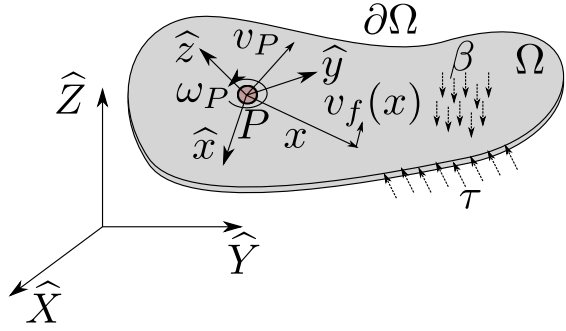} 
	\caption{Thin floating body undergoing a surface traction $\tau$ and body force density $\beta$}
	\label{fig:float_body}
\end{figure}
The coupled ODE-PDE system representing the motion of a single flexible body is here recalled. Then, by exploiting the properties of the cross product, the system of equations is rephrased to highlight the port-Hamiltonian structure.

\subsection{Classical model}

Consider an open connected set $\Omega \subseteq \mathbb{R}^3$, representing a floating flexible body.  The rigid dynamics is located at point $P$, that is not necessarily the center of mass. The velocity of a generic point is expressed by considering a small flexible displacement superimposed to the rigid motion
\[
\bm{v} = \bm{v}_P + \crmat{\bm{\omega}_P} (\bm{x}+\bm{u}_f) + \bm{v}_f,
\]
where $\bm{x}$ is the position vector of the current point, $\bm{v}_P, \bm{\omega}_P$ are the linear and angular velocities of point $P$  and $\bm{v}_f := \dot{\bm{u}}_f$ is the time derivative of the deformation displacement $\bm{u}_f$ (computed in the body frame). These quantities are evaluated in the body reference frame $\widehat{\bm{x}}, \widehat{\bm{y}}, \widehat{\bm{z}}$ (see Fig. \ref{fig:float_body}). The notation $\crmat{\bm{a}}$ (cross map) denotes the skew-symmetric matrix associated to vector $\bm{a}$ (see Appendix A). The model for the classical equations derived using the principle of virtual work can be found in \cite{MB_Daepde} and \cite[Chapter 4]{simeon2013computational}. The small difference with respect to the derivation therein is that the equation for the translation is now written in the body frame (see Appendix B). 
\begin{itemize}
	\item Linear momentum balance:
	\begin{equation}
	\label{eq:rig_tr1}
	\begin{split}
	&m (\dot{\bm{v}}_P + \crmat{\bm{\omega}_P} \bm{v}_P) + \crmat{\bm{s}_u}^\top \dot{\bm{\omega}}_P  + \int_{\Omega} \rho \ddot{\bm{u}}_f \d{\Omega} = \\
	&\quad - \crmat{\bm{\omega}_P} \crmat{\bm{\omega}_P} \bm{s}_u - \int_{\Omega} 2 \rho \crmat{\bm{\omega}_P} \dot{\bm{u}}_f \d{\Omega} +  \int_{\Omega} \bm\beta \d{\Omega} + \int_{\partial \Omega} \bm\tau \d{\Gamma},
	\end{split}
	\end{equation}
	where $\rho$ is the mass density, $m = \int_{\Omega} \rho \d{\Omega}$ the total mass,  $\bm{s}_u = \int_{\Omega} \rho (\bm{x}+\bm{u}_f) \d{\Omega}$ the static moment. Additionally, $\bm\beta$ is a density force and $\bm\tau$ is a surface traction, both expressed in the body reference frame.
	\item Angular momentum balance:
	\begin{equation}
	\label{eq:rig_rot1}
	\begin{split}
	\crmat{\bm{s}_u} (\dot{\bm{v}}_P + \crmat{\bm{\omega}_P} \bm{v}_P) + \bm{J}_u \dot{\bm\omega}_P + \int_{\Omega} \rho \crmat{\bm{x}+\bm{u}_f} \ddot{\bm{u}}_f \d{\Omega} + \crmat{\bm{\omega}_P} \bm{J}_u \bm{\omega}_P = \\ 
	- \int_{\Omega} 2\rho \crmat{\bm{x}+\bm{u}_f} \crmat{\bm\omega_P} \dot{\bm{u}}_f \d{\Omega} + \int_{\Omega}\crmat{\bm{x}+\bm{u}_f} \bm\beta \d{\Omega} + \int_{\partial \Omega}\crmat{\bm{x}+\bm{u}_f} \bm\tau \d{\Gamma}, \\
	\end{split}
	\end{equation}
	where $\bm{J}_u:= \int_{\Omega} \rho \crmat{\bm{x}+\bm{u}_f}^\top\crmat{\bm{x}+\bm{u}_f} \d{\Omega}= - \int_{\Omega} \rho \crmat{\bm{x}+\bm{u}_f}\crmat{\bm{x}+\bm{u}_f} \d{\Omega}$ is the inertia matrix.
	\item Flexibility PDE:
	\begin{equation}
	\label{eq:flex1}
	\begin{split}
	\rho (\dot{\bm{v}}_P + \crmat{\bm\omega_P} \bm{v}_P) + \rho (\crmat{\dot{\bm\omega}_P} + \crmat{\bm{\omega}_P}\crmat{\bm{\omega}_P})(\bm{x}+\bm{u}_f) + \rho (2 \crmat{\bm{\omega}_P} \dot{\bm{u}}_f + \ddot{\bm{u}}_f) = \\
	 \Div{\bm\Sigma} + \bm\beta,
		\end{split}
	\end{equation}
	Variable $\bm\Sigma$ is the Cauchy stress tensor. From linear elasticity theory it is well known that the infinitesimal stress is given by $\bm\varepsilon = \Grad(\bm{u}_f)$, where $\Grad=~\frac{1}{2}[\nabla+\nabla^\top]$ is the symmetric gradient. The constitutive equation is expressed as $\bm\Sigma =  \bm{\mathcal{D}} \bm\varepsilon$, where $ \bm{\mathcal{D}}$ is the stiffness tensor. This PDE requires the specifications of boundary conditions.
	\begin{equation}
	\label{eq:bcPDE}
	\begin{aligned}
	\bm\Sigma \cdot \bm{n}|_{\Gamma_N} &= \bm\tau|_{\Gamma_N}, \quad \text{$\bm{n}$ is the outward normal,} \\
	\bm{u}_f|_{\Gamma_D} &= \bm{\bar{u}}_f|_{\Gamma_D},
	\end{aligned}
	\end{equation}
	The boundary $\partial \Omega = \Gamma_D \cup \Gamma_N$ is split into two subsets, one on which the surface traction is imposed ($\Gamma_N$ Neumann condition) and the other where the flexible displacement is known ($\Gamma_D$ Dirichlet condition). 
\end{itemize}

\subsection{Towards a pH formulation}
The gyroscopic terms in Eqs. \eqref{eq:rig_tr1}, \eqref{eq:rig_rot1}, \eqref{eq:flex1} need some manipulation so that the skew-symmetric interconnection operator can be more easily highlighted. Considering that $\dot{\bm{v}}_f = \ddot{\bm{u}}_f$ and using the Jacobi identity \eqref{eq:jacobi} (see Appendix B for a detailed explanation) the classical equations can be equivalently rewritten as follows. \\
\begin{itemize}
\item Linear momentum balance:
\begin{equation}
\label{eq:rig_tr2}
\begin{split}
m\dot{\bm{v}}_P + \crmat{\bm{s}_u}^\top \dot{\bm{\omega}}_P +   \int_{\Omega} \rho \dot{\bm{v}}_f \d{\Omega}  = \\
\left[m \bm{v}_P + \crmat{\bm{s}_u}^\top \bm\omega_P +2 \int_{\Omega} \rho \bm{v}_f \d{\Omega} \right]_\times \bm\omega_P +  \int_{\Omega} \bm\beta \d{\Omega} + \int_{\partial \Omega} \bm\tau \d{\Gamma}.
\end{split}
\end{equation}
\item Angular momentum balance:
\begin{equation}
\label{eq:rig_rot2}
\begin{split}
\crmat{\bm{s}_u} \dot{\bm{v}}_P  + \bm{J}_u \dot{\bm\omega}_P + \int_{\Omega} \rho \crmat{\bm{x}+\bm{u}_f} \dot{\bm{v}}_f \d{\Omega} = \\
\left[\crmat{\bm{s}_u}^\top \bm\omega_P + 2 \int_{\Omega} \rho \bm{v}_f \d{\Omega} \right]_\times \bm{v}_P + \left[\crmat{\bm{s}_u} \bm{v}_P + \bm{J}_u \bm\omega_P + 2 \int_{\Omega} \rho \crmat{\bm{x}+\bm{u}_f} {\bm{v}}_f \d{\Omega} \right]_\times \bm\omega_P + 
\\
2 \int_{\Omega} \left[\rho \bm{v}_P + \rho \crmat{\bm{x}+\bm{u}_f}^\top \, \bm\omega_P \right]_\times \bm{v}_f \d{\Omega} + \int_{\Omega}\crmat{\bm{x}+\bm{u}_f} \bm\beta \d{\Omega} + \int_{\partial \Omega}\crmat{\bm{x}+\bm{u}_f} \bm{\tau} \d{\Gamma}.
\end{split}
\end{equation}
\item Flexibility PDE:
\begin{equation}
\label{eq:flex2}
\begin{split}
\rho \dot{\bm{v}}_P + \rho \crmat{\bm{x}+\bm{u}_f}^\top \dot{\bm\omega}_P  + \rho \dot{\bm{v}}_f = \\
\left[\rho \bm{v}_P + \rho \crmat{\bm{x}+\bm{u}_f}^\top \bm\omega_P + 2 \rho \bm{v}_f \right]_\times \bm\omega_P + \Div{\bm\Sigma} + \bm\beta.
\end{split}
\end{equation}
Again this equation requires the specification of the boundary conditions \eqref{eq:bcPDE}.
\end{itemize}

 By introducing the appropriate momenta, this model can be reformulated as a pH system as illustrated in the following section.

\section{Elastic body under large rigid motion as a pH system}
\label{sec:pH_fd}
In this section the flexible dynamics of a floating body is written as a coupled system of ODEs and PDEs in pH form. The final form is a descriptor port-Hamiltonian system that fits and generalizes the framework detailed in \cite{beattie2018linear,mehrmann2019structurepreserving}.  

\subsection{Energies and canonical momenta}
Consider the total energy (Hamiltonian), given by the sum of kinetic and deformation energy:
\begin{equation}
\label{eq:H}
\begin{aligned}
H &= H_{\text{kin}} + H_{\text{def}}, \\
&= \frac{1}{2} \int_{\Omega} \left\{\rho ||\bm{v}_P + \crmat{\bm{\omega}_P} (\bm{x}+\bm{u}_f) + {\bm{v}}_f||^2 + \bm\Sigma \cddot \bm\varepsilon \right\}  \d{\Omega}.
\end{aligned}
\end{equation}
The inner product $\bm{A} \cddot \bm{B} = \Tr(\bm{A} \bm{B}^T)$ is the tensor contraction.  
The momenta (usually called energy variables in the pH framework) are then computed by derivation of the Hamiltonian. As the variables belong to finite- and infinite-dimensional spaces the derivative is either a classical gradient or a variational derivative:
\begin{equation}
\label{eq:momenta}
\begin{aligned}
\bm{p}_t &:= \diffp{H}{\bm{v}_P} = m \bm{v}_P + \crmat{\bm{s}_u}^\top \, \bm\omega_P + \int_{\Omega} \rho \bm{v}_f \d{\Omega}, \\
\bm{p}_r &:= \diffp{H}{\bm\omega_P} = \crmat{\bm{s}_u} \bm{v}_P + \bm{J}_u \bm\omega_P + \int_{\Omega} \rho \crmat{\bm{x}+\bm{u}_f} \bm{v}_f \d{\Omega}, \\
\bm{p}_f &:= \diffd{H}{\bm{v}_f} = \rho \bm{v}_P + \rho \crmat{\bm{x}+\bm{u}_f}^\top \, \bm\omega_P + \rho \bm{v}_f, \\
\bm\varepsilon &:= \diffd{H}{\bm\Sigma} = \bm{\mathcal{D}}^{-1} \bm\Sigma,
\end{aligned}
\end{equation}
where the last derivative is computed with respect to a tensor \cite{BRUGNOLI2019940}.
The relation between energy and co-energy variables is then given by
\begin{equation}
\label{eq:mass_op}
\begin{bmatrix}
\bm{p}_t \\ \bm{p}_r \\ \bm{p}_f \\ \bm\varepsilon \\
\end{bmatrix} = 
\underbrace{\begin{bmatrix}
	m \bm{I}_{3\times 3} & \crmat{\bm{s}_u}^\top & \mathcal{I}_\rho^{\Omega} & 0 \\
	\crmat{\bm{s}_u} & \bm{J}_u & \bm{\mathcal{I}}_{\rho x}^{\Omega} & 0  \\
	(\mathcal{I}_\rho^{\Omega})^* & (\bm{\mathcal{I}}_{\rho x}^{\Omega})^* & \rho & 0  \\
	0 & 0 & 0 & \bm{\mathcal{D}}^{-1} \\
	\end{bmatrix}}_{\bm{\mathcal{M}}: \; \text{Mass operator}}
\begin{bmatrix}
\bm{v}_P \\ \bm{\omega}_P  \\ \bm{v}_f  \\ \bm\Sigma \\
\end{bmatrix},
\end{equation}
where $\bm{I}_{3\times 3}$ is the identity matrix in $\mathbb{R}^3$. The operators are defined as
\begin{equation*}
\begin{aligned}
\mathcal{I}_\rho^\Omega &:=\int_{\Omega} \rho (\cdot) \d{\Omega}, \\
(\mathcal{I}_\rho^{\Omega})^* &= \rho, \\
\end{aligned} \qquad
\begin{aligned} 
\bm{\mathcal{I}}_{\rho x}^{\Omega} &:= \int_\Omega \rho \crmat{\bm{x}+\bm{u}_f} (\cdot) \d{\Omega}, \\
(\bm{\mathcal{I}}_{\rho x}^{\Omega})^* &= \rho \crmat{\bm{x}+\bm{u}_f}^\top= -\rho \crmat{\bm{x}+\bm{u}_f}. \\
\end{aligned}
\end{equation*}
The superscript $*$ denotes the adjoint operator (see Appendix A). The mass operator $\bm{\mathcal{M}}$ is a self-adjoint, positive operator. The kinetic and deformation energy can then be written as
\begin{equation}
H_{\text{kin}} + H_{\text{def}} = \frac{1}{2} \langle \bm{e}_{\text{kd}}, \ \bm{\mathcal{M}} \bm{e}_{\text{kd}} \rangle
\end{equation}
where $\bm{e}_{\text{kd}} = [\bm{v}_P; \, \bm{\omega}_P; \, \bm{v}_f; \bm{\Sigma}]$ and the inner product $\langle \ , \ \rangle$ is taken over the space $\mathbb{R}^3 \times \mathbb{R}^3 \times \mathscr{L}^2(\Omega, \mathbb{R}^3) \times \mathscr{L}^2(\Omega, \mathbb{R}^{3\times 3}_{\text{sym}})$ ($\mathscr{L}^2(\Omega, \mathbb{R}^3), \; \mathscr{L}^2(\Omega, \mathbb{R}^{3\times 3}_{\text{sym}})$ are the spaces of square integrable vector-valued or symmetric tensors valued functions in $\mathbb{R}^3$). Notice that the kinetic energy also depends on the flexible displacement
\[
\diffd{H_{\text{kin}}}{\bm{u}_f} = \crmat{\bm{p}_f} \bm{\omega_{P}}.
\]
This term is responsible for a coupling between the kinematic coordinates and the velocities, as will be clear in the following section. 

\subsection{PH formulation}
In order to get a complete formulation, generalized coordinates are required. It is natural to select the following variables:
\begin{itemize}
	\item $^i \bm{r}_P$ the position of point $P$ in the inertial frame of reference;
	\item $\bm{R}$ the direction cosine matrix that transforms vectors from the body frame to the inertial frame (other attitude parametrizations are possible, here the direction cosine matrix is considered for ease of presentation);
	\item $\bm{u}_f$ the flexible displacement;
\end{itemize}

In particular, following \cite{attitude_ph}, the direction cosine matrix is converted into a vector by concatenating its rows
\begin{equation*}
\bm{R}_{\text{v}} = \text{vec}(\bm{R}^\top) = [\bm{R}_x \; \bm{R}_y \; \bm{R}_z]^\top,
\end{equation*}
where $\bm{R}_{x}, \bm{R}_{y}, \bm{R}_{z}$ are the first, second and third row of matrix $\bm{R}$. Furthermore the corresponding cross map will be given by
\begin{equation*}
\crmat{\bm{R}_{\text{v}}} = 
\begin{bmatrix}
\crmat{\bm{R}_x} \\
\crmat{\bm{R}_y} \\
\crmat{\bm{R}_z} \\
\end{bmatrix}, \qquad 
\crmat{\bm{R}_{\text{v}}} : \mathbb{R}^9 \rightarrow \mathbb{R}^{9 \times 3}.
\end{equation*}

The overall port-Hamiltonian formulation, equivalent to Eqs. \eqref{eq:rig_tr2}, \eqref{eq:rig_rot2}, \eqref{eq:flex2}, is then (omitting the external forces and torques)
\begin{equation}
\label{eq:ph_mfd_all}
\setlength{\dashlinegap}{2pt}
\underbrace{
	{\left[ \begin{array}{c:c}
		\bm{I} & 0 \\
		\hdashline
		0 & \bm{\mathcal{M}} \\
		\end{array} \right]}
}_{\bm{\mathcal{E}}}
\diff{}{t}
\underbrace{\begin{bmatrix}
	^i \mathbf{r}_P \\ \bm{R}_{\text{v}} \\ \bm{u}_f \\\hdashline  \bm{v}_P \\ \bm\omega_P  \\ \bm{v}_f  \\ \bm\Sigma \\
	\end{bmatrix}}_{\bm{e}} = 
\underbrace{
	{\left[ \begin{array}{ccc:cccc}
		0 & 0 & 0 &  \bm{R} & 0 & 0 & 0 \\
		0 & 0 & 0 & 0 & \crmat{\bm{R}_{\text{v}}} & 0 & 0 \\
		0 & 0 & 0 & 0 & 0 & \bm{I}_{3\times 3} & 0  \\ 
		\hdashline
		-\bm{R}^\top & 0 & 0 & 0 & \crmat{\widetilde{\bm{p}}_t} & 0 & 0 \\
		0 & -\crmat{\bm{R}_{\text{v}}}^\top & 0 & \crmat{\widetilde{\bm{p}}_t} & \crmat{\widetilde{\bm{p}}_r} & \bm{\mathcal{I}}_{p_f}^\Omega & 0 \\
		0 & 0 & -\bm{I}_{3\times 3} & 0 & -(\bm{\mathcal{I}}_{p_f}^\Omega)^* & 0 & \Div \\
		0 & 0 & 0 & 0 & 0 & \Grad & 0 \\
		\end{array} \right]}
}_{\bm{\mathcal{J}}}
\underbrace{\begin{bmatrix}
	\partial_{\bm{r}_P}H \\ \partial_{\bm{R}_\text{v}}H \\ \delta_{\bm{u}_f} H \\\hdashline  \bm{v}_P \\ \bm\omega_P  \\ \bm{v}_f  \\ \bm\Sigma \\
	\end{bmatrix}}_{\bm{z}}.
\end{equation} 

Variables $\widetilde{\bm{p}}_t, \widetilde{\bm{p}}_r$ are defined as
\begin{equation}
\label{eq:mod_momenta}
\begin{aligned}
\widetilde{\bm{p}}_t &= \bm{p}_t + \int_{\Omega} \rho \bm{v}_f \d{\Omega}, \\
\widetilde{\bm{p}}_r &= \bm{p}_r + \int_{\Omega} \rho \crmat{\bm{x} + \bm{u}_f}\bm{v}_f \d{\Omega}. \\
\end{aligned}
\end{equation}
The operator $\bm{\mathcal{I}}_{p_f}^\Omega$ is defined as 
\begin{equation}
\label{eq:mod_momentafl}
\bm{\mathcal{I}}_{p_f}^\Omega := \int_\Omega \left\{2 \crmat{\bm{p}_f} + \rho \crmat{\bm{v}_f} \right\}(\cdot) \d{\Omega}.
\end{equation}
Its  adjoint is given by
\begin{equation*}
(\bm{\mathcal{I}}_{p_f}^\Omega)^* = \left\{2 \crmat{\bm{p}_f}^\top + \rho \crmat{\bm{v}_f}^\top \right\}(\cdot) = - \left\{2 \crmat{\bm{p}_f} + \rho \crmat{\bm{v}_f} \right\}(\cdot).
\end{equation*} 
The coefficient 2 is required to compensate the contribution given by $\delta_{\bm{u}_f} H$ 
\[
-\diffd{H}{\bm{u}_f} - (\bm{\mathcal{I}}_{p_f}^\Omega)^* \bm{\omega}_P = \left[\rho \bm{v}_P + \rho \crmat{\bm{x}+\bm{u}_f}^\top \bm\omega_P + 2 \rho \bm{v}_f \right]_\times \bm\omega_P.
\]
The additional terms related to $\rho \bm{v}_f$ are associated to the Coriolis accelerations that affect the deformation field. It is important to underline that $\Div$ and $\Grad$ are formally skew-adjoint operators, i.e. for homogeneous boundary conditions (I.B.P. stands for integration by parts)
\begin{align*}
\int_{\Omega} \bm\Sigma \cddot \Grad(\bm{v}_f) \d{\Omega} &\underbrace{=}_{\text{I.B.P.}} -\int_{\Omega} \Div(\bm\Sigma) \cdot \bm{v}_f \d{\Omega}, \\
\left\langle \bm\Sigma, \, \Grad(\bm{v}_f) \right\rangle_{\mathscr{L}^2(\Omega, \mathbb{R}^{3\times 3}_{\text{sym}})} &\underbrace{=}_{\text{I.B.P.}} -\left\langle\Div(\bm\Sigma), \, \bm{v}_f \right\rangle_{\mathscr{L}^2(\Omega, \mathbb{R}^3)}, 
\end{align*}
where $\left\langle ,  \right\rangle_\mathscr{H}$ denote an inner product over the Hilbert space $\mathscr{H}$. For this reason the operator $\bm{\mathcal{J}}$ is skew-symmetric $\bm{\mathcal{J}}_{}^*=-\bm{\mathcal{J}}$. System \eqref{eq:ph_mfd_all} fits into the framework detailed in \cite{mehrmann2019structurepreserving} and extends it, since a coupled system of ODEs and PDEs is considered. The underlying state space is 
\[
\mathscr{X} = \mathbb{R}^3 \times \mathbb{R}^9 \times \mathscr{L}^2(\Omega, \mathbb{R}^{3}) \times \mathbb{R}^3 \times \mathbb{R}^3 \times \mathscr{L}^2(\Omega, \mathbb{R}^{3}) \times \mathscr{L}^2(\Omega, \mathbb{R}^{3\times 3}_{\text{sym}}).
\] 
The dynamics can be rewritten compactly as follows
\begin{equation}
\label{eq:MFD_pHDAE}
\begin{aligned}
\bm{\mathcal{E}}(\bm{e}) \diffp{\bm{e}}{t} &= \bm{\mathcal{J}}(\bm{e}) \bm{z}(\bm{e}) + \bm{\mathcal{B}}_d(\bm{e}) \bm{u}_d + \bm{\mathcal{B}}_r(\bm{e}) \bm{u}_\partial, \\
\bm{y}_d &= \bm{\mathcal{B}}_d^*(\bm{e}) \bm{z}(\bm{e}), \\
\bm{y}_r &= \bm{\mathcal{B}}_r^*(\bm{e}) \bm{z}(\bm{e}), \\
\bm{u}_\partial &= \bm{\mathcal{B}}_{\partial} \bm{z}(\bm{e}) =  \bm\Sigma \cdot \bm{n}|_{\partial \Omega} = \bm\tau|_{\partial \Omega}, \\
\bm{y}_\partial &= \bm{\mathcal{C}}_{\partial} \bm{z}(\bm{e}) = \bm{v}_f|_{\partial \Omega},
\end{aligned}
\end{equation}
where $\bm{u}_d = \bm\beta$. Using definitions  \eqref{eq:momenta}, it follows that the Hamiltonian  satisfies 
\begin{equation}
\label{eq:gradH}
\partial_{\bm{e}} H = \bm{\mathcal{E}}^* \bm{z}.
\end{equation}

Adopting the same nomenclature as in \cite{mehrmann2019structurepreserving}, $\bm{e}$ contains the state and $\bm{z}$ contains the effort functions. The operators verify $\bm{\mathcal{E}} = \bm{\mathcal{E}}^*, \; \bm{\mathcal{J}} = -\bm{\mathcal{J}}^*$. The control operators are expressed as
\begin{align*}
\bm{\mathcal{B}}_d &= 
\begin{bmatrix}
0 & 0 & 0 & \mathcal{I}^\Omega & \bm{\mathcal{I}}_{x}^\Omega & \bm{I} & 0
\end{bmatrix}^\top, \\
\bm{\mathcal{B}}_r &= 
\begin{bmatrix}
0 & 0 & 0 & \mathcal{I}^{\Gamma} & \bm{\mathcal{I}}_{x}^{\Gamma} & 0 & 0 \\
\end{bmatrix}^\top,
\end{align*}
where 
\begin{equation*}
\begin{aligned}
\mathcal{I}^\Omega &:=\int_{\Omega} (\cdot) \d{\Omega}, \\
\mathcal{I}^{\Gamma} &:= \int_{\partial \Omega} (\cdot) \d{\Gamma}, \\
\end{aligned} \qquad
\begin{aligned} 
\bm{\mathcal{I}}_{x}^\Omega &:=\int_{\Omega} \crmat{\bm{x}+\bm{u}_f} (\cdot) \d{\Omega}, \\
\bm{\mathcal{I}}_{x}^{\Gamma} &:=\int_{\partial \Omega} \crmat{\bm{x}+\bm{u}_f} (\cdot) \d{\Gamma}. \\
\end{aligned}
\end{equation*}
The distributed control operator $\bm{\mathcal{B}}_d$  is compact. The boundary traction force acts on the rigid part through the compact operator $\bm{\mathcal{B}}_r$. Notice that by definition of adjoint (see Appendix A), the vector $\bm{y}_r$ represents the rigid body velocity at the boundary
\[
\bm{y}_r = (\bm{v}_P + \crmat{\bm{x}+\bm{u}_f}^\top \bm{\omega}_P)\vert_{\partial\Omega},
\] 
while $\bm{y}_d$ represents the velocity field in the domain
\[
\bm{y}_d = (\bm{v}_P + \crmat{\bm{x}+\bm{u}_f}^\top \bm{\omega}_P + \bm{v}_f)\vert_{\Omega}.
\]
The power balance is naturally embedded in the dynamics 
\begin{equation}
\begin{aligned}
\dot{H}(\bm{e}) &= \langle \partial_{\bm{e}} H, \partial_t {\bm{e}} \rangle_{\mathscr{X}} = \langle \bm{\mathcal{E}}^* \bm{z}, \partial_t {\bm{e}} \rangle_{\mathscr{X}}, \\
&= \langle \bm{z}, \bm{\mathcal{E}} \partial_t {\bm{e}} \rangle_{\mathscr{X}}, \quad \text{Adjoint definition}, \\
& = \langle \bm{z}, \bm{\mathcal{J}}\bm{z} + \bm{\mathcal{B}}_d(\bm{e}) \bm{u}_d + \bm{\mathcal{B}}_r(\bm{e}) \bm{u}_\partial \rangle_{\mathscr{X}}, \\
& = \langle \bm{y}_\partial,  \bm{u}_\partial \rangle_{\mathscr{L}^2(\partial\Omega, \mathbb{R}^3)} + \langle \bm{\mathcal{B}}_d^* \bm{z}, \bm{u}_d \rangle_{\mathscr{X}} + \langle \bm{\mathcal{B}}_r^* \bm{z}, \bm{u}_\partial \rangle_{\mathscr{X}}, \quad \text{I.B.P. on } \bm{\mathcal{J}}, \\
&= \langle \bm{y}_\partial + \bm{y}_r,  \bm{u}_\partial \rangle_{\mathscr{L}^2(\partial\Omega, \mathbb{R}^3)} + \langle \bm{y}_d,  \bm{u}_d \rangle_{\mathscr{L}^2(\Omega, \mathbb{R}^3)}, \\
\end{aligned}
\end{equation}
where the integration by parts (Stokes theorem) has been used
\begin{equation}
\label{eq:stokes}
\int_{\Omega} \bm\Sigma \cddot \Grad(\bm{v}_f) \d{\Omega} + \int_{\Omega} \Div(\bm\Sigma) \cdot \bm{v}_f \d{\Omega} = \int_{\partial \Omega} (\bm\Sigma \cdot \bm{n}) \cdot \bm{v}_f \d{\Gamma} = \langle \bm{y}_\partial,  \bm{u}_\partial \rangle_{\mathscr{L}^2(\partial\Omega)}.
\end{equation}
The power balance equals the power due to body force and surface traction
\begin{equation}
\dot{H}(\bm{e}) = \int_{\partial \Omega} (\bm{\Sigma} \cdot \bm{n}) \cdot \bm{v} \d\Gamma + \int_{\Omega} \bm{u}_d \cdot \bm{v}  \d{\Omega}, \quad \bm{v} := \bm{v}_P + \crmat{\bm{\omega}_P} (\bm{x}+\bm{u}_f) + {\bm{v}}_f.
\end{equation}
Even if three dimensional elasticity has been taken as example up to this point, other models are easily considered. Beam and plate models \cite{BRUGNOLI2019940,BRUGNOLI2019961} are described by appropriate differential operators that replace the $\Div, \Grad$ appearing in \eqref{eq:ph_mfd_all} (see \secref{sec:ph_floatbeam}).

\begin{remark}
	Conservative forces are easily accounted for by introducing an appropriate potential energy. For example if the gravity force is considered, the corresponding potential energy reads
	\begin{equation*}
	H_{\text{pot}} = \int_{\Omega} \rho g \, ^i r_z \d{\Omega} = \int_{\Omega} \rho g \left[ ^i r_{P, z} +\bm{R}_z (\bm{x} + \bm{u}_f) \right] \d{\Omega},
	\end{equation*}
	where $^i r_z$ is the vertical location of a generic point computed in the inertial frame. The associated co-energy variables are easily obtained
	\begin{align*}
	\partial_{\bm{r}_P}H_{\text{pot}} &= m g \, \widehat{\bm{Z}}, \quad \text{$\widehat{\bm{Z}}$ is the inertial frame vertical direction},\\
	\partial_{\bm{R}_\text{v}}H_{\text{pot}} &= [\bm{0}_{(3, 1)}, \; \bm{0}_{(3, 1)}, \; \int_{\Omega} \rho g (\bm{x} + \bm{u}_f)^\top \d{\Omega}]^\top, \\
	\delta_{\bm{u}_f} H_{\text{pot}} &= \rho g \, \bm{R}_z^\top.
	\end{align*}
	These contributions correspond to the forcing terms due to gravity.
\end{remark}

\begin{remark}
	The linear elasticity hypothesis does not allow including the effect of non-linearities due to large deformations.  However, geometric stiffening could be considered by adding a potential energy associated to centrifugal forces \cite{MB_Daepde}. 
\end{remark}

\begin{remark}
	If case of vanishing deformations $\bm{u}_f \equiv 0$, the Newton-Euler equations on the Euclidean group $SE(3)$ are retrieved \cite{celledoni2018passivity}
	\begin{equation*}
	\begin{bmatrix}
	^i\dot{\bm{r}}_P \\ \bm{R}_{\text{v}} \\\dot{\bm{p}}_t \\ \dot{\bm{p}}_r \\
	\end{bmatrix} = 
	\begin{bmatrix}
	0 & 0 & \bm{R} & 0 \\
	0 & 0 & 0 & \crmat{\bm{R}_{\text{v}}} \\
	- \bm{R}^\top & 0 & 0 & \crmat{\bm{p}_t}\\
	0 & -\crmat{\bm{R}_{\text{v}}}^\top & \crmat{\bm{p}_t} & \crmat{\bm{p}_r} \\
	\end{bmatrix}
	\begin{bmatrix}
	\partial_{\bm{r}_P} H \\ \partial_{\bm{R}_{\text{v}}} H \\ \bm{v}_P \\ \bm{\omega}_P  \\
	\end{bmatrix},
	\end{equation*}
	where
	\begin{equation*}
	\begin{bmatrix}
	\bm{p}_t \\ \bm{p}_r \\ 
	\end{bmatrix} = 
	\begin{bmatrix}
	m \bm{I} & \crmat{\bm{s}}^\top \\
	\crmat{\bm{s}} & \bm{J} \\
	\end{bmatrix}
	\begin{bmatrix}
	\bm{v}_P \\ \bm{\omega}_P  \\ 
	\end{bmatrix}, \qquad \bm{p} = \bm{M} \bm{v}.
	\end{equation*}
	The kinetic energy is then given by $H_{\text{kin}} = \frac{1}{2} \bm{v}^\top \bm{M} \bm{v}$.
	This system can be written in standard pH form as $\dot{\bm{x}} = \bm{J}(\bm{x})\partial_{\bm{x}} H$.
\end{remark}

\section{Discretization procedure}
\label{sec:discr}
A finite-element based technique to obtain a finite-dimensional pH system is illustrated. This methodology relies on the results explained in \cite{cardoso2019partitioned} and ahead, used in \cite{BRUGNOLI2019940,BRUGNOLI2019961}. The essential feature of this method is that it is structure-preserving. Given the lossless infinite-dimensional system \eqref{eq:MFD_pHDAE}, it allows obtaining a finite-dimensional representation that is again lossless. The procedure boils down to three simple steps
\begin{enumerate}
	\item The system is written in weak form; 
	\item An integration by parts is applied to highlight the appropriate boundary control;
	\item A Galerkin method is employed to obtain a finite-dimensional system.
\end{enumerate}

\subsection{Illustration for the Elastodynamics PDE}
To explain the methodology, consider the elastodynamics PDE
\begin{equation*}
\rho \diffp[2]{\bm{u}}{t} - \Div\left(\bm{\mathcal{D}} \Grad(\bm{u})\right) = \bm{u}_d,
\end{equation*}
where a distributed control $\bm{u}_d$ (a volumetric force) is considered. This model describes the flexible vibrations of a continuum under small deformations. It is embedded in the general formulation \eqref{eq:ph_mfd_all} and therefore the procedure explained here is easily adapted to the general formulation. \\
To get a pH representation, the energy variables have to be properly selected by considering the total energy
\begin{equation}
\label{eq:en_eldyn}
\displaystyle H = \frac{1}{2} \int_{\Omega} \left\{\rho \left(\diffp{\bm{u}}{t}\right)^2 + \bm\Sigma \cddot \bm\varepsilon \right\} \d{\Omega}.
\end{equation}
Taking as energy variables the linear momentum and the deformation
\begin{equation}
\begin{aligned}
\text{Energies} \quad  \bm{x}_1 &:= \rho \ \partial_t \bm{u}, \\
\text{Co-energies} \quad \bm{e}_1 &:= \diffd{H}{\bm{x}_1} =  \partial_t \bm{u}, \\
\end{aligned} 
\end{equation}
the corresponding co-energies are obtained by taking the variational derivative of the Hamiltonian
\begin{equation}
\begin{aligned}
\bm{X}_2 &:= \bm\varepsilon = \Grad(\bm{u}). \\
\bm{E}_2 &:= \diffd{H}{\bm{X}_2} = \bm\Sigma.
\end{aligned}
\end{equation}
The port-Hamiltonian representation in co-energy variables becomes
\begin{equation*}
\underbrace{\begin{bmatrix}
	\rho & 0 \\ 0 & \bm{\mathcal{D}}^{-1} \\
	\end{bmatrix}}_{\bm{\mathcal{M}}}
\diffp{}{t}
\begin{bmatrix}
\bm{e}_1 \\ \bm{E}_2 \\
\end{bmatrix} = 
\underbrace{\begin{bmatrix}
	0 & \Div \\ \Grad & 0 \\
	\end{bmatrix}}_{\bm{\mathcal{J}}}
\begin{bmatrix}
\bm{e}_1 \\ \bm{E}_2 \\
\end{bmatrix} + 
\underbrace{\begin{bmatrix}
	\bm{I} \\ 0 \\
	\end{bmatrix}}_{\bm{\mathcal{B}}_d} \bm{u}_d 
\end{equation*}
The interconnection operator may be decomposed as $\bm{\mathcal{J}} = \bm{\mathcal{J}}_{\Div} + \bm{\mathcal{J}}_{\Grad}$
\begin{equation}
\underbrace{\begin{bmatrix}
	0 & \Div \\ \Grad & 0 \\
	\end{bmatrix}}_{\bm{\mathcal{J}}} = 
\underbrace{\begin{bmatrix}
	0 & \Div \\ 0  & 0 \\
	\end{bmatrix}}_{\bm{\mathcal{J}}_{\Div}} + 
\underbrace{\begin{bmatrix}
	0 & 0 \\ \Grad & 0 \\
	\end{bmatrix}}_{\bm{\mathcal{J}}_{\Grad}}
\end{equation}
Assuming a Neumann boundary conditions (the normal traction $\bm\tau$ is known at the boundary), this system can be written compactly as a boundary control system
\begin{equation}
\label{eq:eldyn_PDE}
\begin{aligned}
\bm{\mathcal{M}} \diffp{\bm{e}}{t} &= \bm{\mathcal{J}} \bm{e} + \bm{\mathcal{B}}_d \bm{u}_d, \\
\bm{y}_d &= \bm{\mathcal{B}}_d^* \bm{e}, \\
\bm{u}_\partial &= \bm{E}_2 \cdot \bm{n}|_{\partial\Omega}, \\
\bm{y}_\partial &= \bm{e}_1|_{\partial\Omega}.
\end{aligned}
\end{equation}
The system is defined over the state space
\[
\mathscr{X} = \mathscr{L}^2(\Omega, \mathbb{R}^3) \times\mathscr{L}^2(\Omega, \mathbb{R}^{3\times 3}_{\text{sym}}),
\]
where $\mathscr{L}^2$ is the space of square integrable functions. Taking  $[\bm{a}, \bm{A}], \ [\bm{b}, \bm{B}] \in \mathscr{X}$ the inner product is computed as
\[
\left\langle [\bm{a}, \bm{A}], \ [\bm{b}, \bm{B}] \right\rangle_{\mathscr{X}} = \int_{\Omega} \bm{a} \cdot \bm{b} \d{\Omega} + \int_{\Omega} \bm{A} \cddot \bm{B} \d{\Omega}.
\]
The total energy is then computed as an inner product modulated by the mass operator $H = \frac{1}{2} \left\langle \bm{e}, \ \bm{\mathcal{M}} \bm{e} \right\rangle_{\mathscr{X}}$ (see \eqref{eq:en_eldyn}). The power balance is computed by applying the Stokes theorem \eqref{eq:stokes}
\begin{equation}
\label{eq:pow_eldyn}
\dot{H} = \left\langle \bm{e},  \bm{\mathcal{M}} \partial_t \bm{e} \right\rangle_{\mathscr{X}} = \langle \bm{y}_\partial,  \bm{u}_\partial \rangle_{\mathscr{L}^2(\partial\Omega, \mathbb{R}^3)} + \langle \bm{y}_d,  \bm{u}_d \rangle_{\mathscr{L}^2(\Omega, \mathbb{R}^3)}.
\end{equation}
So the system is lossless and passive with storage function given by the total energy. Considering a test function $\bm{w} = [\bm{w}_1, \; \bm{W}_2]$ the weak form reads
\begin{equation*}
\left\langle \bm{w}, \; \bm{\mathcal{M}} \ \partial_t \bm{e} \right\rangle_{\mathscr{X}} = \left\langle \bm{w}, \; \bm{\mathcal{J}} \bm{e} \right\rangle_{\mathscr{X}} + \left\langle \bm{w}, \; \bm{\mathcal{B}}_d \bm{u}_d \right\rangle_{\mathscr{X}}.
\end{equation*}
The bilinear form $m(\bm{w}, \partial_t \bm{e}) = \left\langle \bm{w}, \; \bm{\mathcal{M}} \ \partial_t \bm{e} \right\rangle_{\mathscr{X}}$ is symmetric and coercive. The bilinear form $b_d(\bm{w}, \bm{u}_d):=\left\langle \bm{w}, \; \bm{\mathcal{B}}_d \bm{u}_d \right\rangle_{\mathscr{X}}$ takes into account distributed control.\\
Now an integration by parts is applied on $\bm{\mathcal{J}}_{\Div}$:
\begin{equation}
\left\langle \bm{w}, \; \bm{\mathcal{J}} \bm{e} \right\rangle_{\mathscr{X}} = \left\langle \bm{w}, \; \bm{\mathcal{J}}_{\Grad} \bm{e} \right\rangle_{\mathscr{X}} - \left\langle \bm{\mathcal{J}}_{\Grad} \bm{w}, \; \bm{e} \right\rangle_{\mathscr{X}} + \left\langle \bm{w}, \; \bm{u}_\partial \right\rangle_{\mathscr{L}^2(\partial \Omega, \mathbb{R}^3)},
\end{equation}
where $\left\langle \cdot,  \cdot \right\rangle_{\mathscr{L}^2(\partial \Omega, \mathbb{R}^3)}$ denotes the $\mathscr{L}^2$ inner product over the boundary. The expression $j_{\Grad}(\bm{w}, \bm{e}) :=\left\langle \bm{w}, \; \bm{\mathcal{J}}_{\Grad} \bm{e} \right\rangle_{\mathscr{X}} - \left\langle \bm{\mathcal{J}}_{\Grad} \bm{w}, \; \bm{e} \right\rangle_{\mathscr{X}}$ is a skew symmetric bilinear form, since $j_{\Grad}(\bm{w}, \bm{e})=-j_{\Grad}(\bm{e},\bm{w})$ holds. The bilinear form $b_{\partial}(\bm{w}, \bm{u}_\partial) := \left\langle \bm{w}, \; \bm{u}_\partial \right\rangle_{\mathscr{L}^2(\partial \Omega, \mathbb{R}^3)}$ imposes the Neumann condition weakly. System \eqref{eq:eldyn_PDE} is now rewritten in weak form
\begin{equation}
m(\bm{w}, \partial_t \bm{e}) = j_{\Grad}(\bm{w}, \bm{e}) + b_d(\bm{w}, \bm{u}_d) + b_{\partial}(\bm{w}, \bm{u}_\partial).
\end{equation}
The output equation is discretized considering test function $\bm{w}_\partial$ defined over the boundary
\begin{equation}
\left\langle \bm{w}_\partial, \; \bm{y}_\partial \right\rangle_{\mathscr{L}^2(\partial \Omega, \mathbb{R}^3)} = \left\langle \bm{w}_\partial, \; \bm{e}_1 \right\rangle_{\mathscr{L}^2(\partial \Omega, \mathbb{R}^3)}.
\end{equation}
If a Galerkin method is applied then corresponding test and trial functions are discretized using the same basis
\begin{equation*}
\begin{aligned}
\bm{w}_1(\bm{x},t) = \bm{\phi}_1(\bm{x})^\top \mathbf{w}_1(t), \\
\bm{e}_1(\bm{x},t) = \bm{\phi}_1(\bm{x})^\top \mathbf{e}_1(t), 
\end{aligned} \qquad
\begin{aligned}
\bm{W}_2(\bm{x},t) = \bm{\phi}_2(\bm{x})^\top \mathbf{w}_2(t), \\
\bm{E}_2(\bm{x},t) = \bm{\phi}_2(\bm{x})^\top \mathbf{e}_2(t),
\end{aligned}
\end{equation*}
where the bold italic variables represent numerical vectors. A finite-dimensional pH system is readily obtained
\begin{equation}
\begin{aligned}
\mathbf{M} \dot{\mathbf{e}} &= \mathbf{J} \mathbf{e} + \mathbf{B}_d \mathbf{u}_d + \mathbf{B}_\partial \mathbf{u}_\partial, \\
\mathbf{y}_d &:= \mathbf{M}_d \widetilde{\mathbf{y}}_d = \mathbf{B}_d^\top \mathbf{e},  \\
\mathbf{y}_\partial &:= \mathbf{M}_\partial \widetilde{\mathbf{y}}_\partial = \mathbf{B}_\partial^\top \mathbf{e}.
\end{aligned}
\end{equation}
It is important to notice that this system is again lossless. The discrete energy is $H_d = \frac{1}{2} \mathbf{e}^\top \mathbf{M} \mathbf{e}$.
The discrete power balance is given by
\begin{equation*}
\dot{H}_d = \mathbf{e}^\top \mathbf{M} \dot{\mathbf{e}} = \mathbf{e}^\top ( \mathbf{J} \dot{\mathbf{e}} + \mathbf{B}_d \mathbf{u}_d + \mathbf{B}_\partial \mathbf{u}_\partial) = \mathbf{y}_d^\top \mathbf{u}_d + \mathbf{y}_\partial^\top \mathbf{u}_\partial,
\end{equation*}
which mimics \eqref{eq:pow_eldyn} at the discrete level.
\begin{remark}
	Vectors $\widetilde{\mathbf{y}}_d, \widetilde{\mathbf{y}}_\partial$ correspond to the output degrees of freedom. The outputs $\mathbf{y}_d, \mathbf{y}_\partial$ have been defined incorporating the mass matrix in order get the discrete power balance $\dot{H}_d = \mathbf{u}_\partial^\top \mathbf{y}_\partial + \mathbf{u}_d^\top \mathbf{y}_d$.
\end{remark}
\begin{remark}
	Stable mixed finite elements for the elastodynamics problem are detailed in \cite{ArnoldElasDyn}.  The formulation therein is based on a weak form obtained by integration by parts of the $\bm{\mathcal{J}}_{\Grad}$ operator. The mixed finite element method for such a problem are then stable in the sense of Brezzi thanks to the properties of $L^2 / H^{\Div}$ finite element spaces. However, the discretization scheme proposed here allows for an easier representation of floating bodies as the free condition corresponds to zero Neumann boundary conditions.
\end{remark}

\subsection{Discretized rigid-flexible port-Hamiltonian dynamics}

The same methodology is applied to system \eqref{eq:MFD_pHDAE}. If corresponding test functions $w$, state $e$ and effort functions $z$ are discretized using the same bases
\[ \bm{w}(\bm{x}, t) = \bm{\phi}(\bm{x})^\top \mathbf{w}(t), \quad \bm{e}(\bm{x}, t) = \bm{\phi}(\bm{x})^\top \mathbf{e}(t), \quad \bm{z}(\bm{x}, t) = \bm{\phi}(\bm{x})^\top \mathbf{z}(t),
\]
then a finite-dimensional pHDAE system is obtained (after integration by parts of the $\mathcal{J}_{\Div}$ operator)
\begin{equation}
\begin{aligned}
\mathbf{E}(\mathbf{e}) \dot{\mathbf{e}} &= \mathbf{J}(\mathbf{e}) \mathbf{z}(\mathbf{e}) + \mathbf{B}_d(\mathbf{e}) \mathbf{u}_d + \mathbf{B}_\partial(\mathbf{e}) \mathbf{u}_\partial, \\
\mathbf{y}_d &:= \mathbf{M}_d \widetilde{\mathbf{y}}_d = \mathbf{B}_d^\top \mathbf{z}(\mathbf{e}),  \\
\mathbf{y}_\partial &:= \mathbf{M}_\partial \widetilde{\mathbf{y}}_\partial = \mathbf{B}_\partial^\top \mathbf{z}(\mathbf{e}).
\end{aligned}
\end{equation}
The computation of vector $\mathbf{z}$ is based on the discrete Hamiltonian  gradient:
\[
\diffp{H_d}{\mathbf{e}} = \mathbf{E}^\top \mathbf{z}, \qquad H_d = H_{d, \text{kin}}+H_{d, \text{def}}+H_{d, \text{pot}}.
\]
This relation represents the finite-dimensional counterpart of \eqref{eq:gradH}. For the deformation and kinetic energy, it is straightforward to find the link between the state and effort functions since those energies are quadratic in the state variable:
\begin{equation}
H_{d, \text{kin}} + H_{d, \text{def}} = \frac{1}{2} \mathbf{e}_{\text{kd}}^\top \, \mathbf{M}_{\text{kd}} \, \mathbf{e}_{\text{kd}} \longrightarrow \mathbf{z}_{\text{kd}} = \mathbf{e}_{\text{kd}},
\end{equation}
where $\mathbf{e}_{\text{kd}} = [\mathbf{v}_P; \, \bm{\omega}_P; \, \mathbf{v}_f; \bm{\Sigma}]$ and $\mathbf{M}_{\text{kd}}$ is the discretization of the mass operator $\bm{\mathcal{M}}$ given in Eq \eqref{eq:mass_op}.
The only term that requires additional care is the potential energy and particularly the variational derivative of the Hamiltonian with respect to the deformation displacement $\bm{z}_{u}=\delta_{\bm{u}_f} H$.  Consider the continuous power balance associated to the flexible displacement
\[
\dot{H}_u = \int_{\Omega} \diffp{\bm{u}_f}{t} \cdot \bm{z}_{u} \d{\Omega} = \int_{\Omega} \diffp{\bm{u}_f}{t} \cdot \diffd{H}{\bm{u}_f} \d{\Omega}
\]
The deformation velocity and its corresponding effort variable are discretized using the same basis, i.e. $\bm{u}_f = \bm{\phi}_u^\top \mathbf{u}_f, \; \bm{z}_u = \bm{\phi}_u^\top \mathbf{z}_u$. The discrete Hamiltonian rate assumes two equivalent expressions
\begin{equation*}
\dot{H}_{u, d}(\mathbf{u}_f) = 
\begin{cases}
\dot{\mathbf{u}}_f^\top \mathbf{M}_u \; \mathbf{z}_u, \\
\displaystyle \dot{\mathbf{u}}_f^\top \diffp{H_d}{\mathbf{u}_f},
\end{cases}
\end{equation*}
where $\mathbf{M}_u=\int_{\Omega} \bm{\phi}_u \, \bm{\phi}_u^\top \d{\Omega}$. To preserve the power balance at the discrete level, $ \mathbf{z}_u = \mathbf{M}_u^{-1} \diffp{H_d}{\mathbf{u}_f}$ must hold. \\

\begin{remark}\label{rmk:dirich}
	The set $\Gamma_D$ for the Dirichlet condition has to be non empty, otherwise the deformation field is allowed for rigid movement, leading to a singular mass matrix. To enforce that, test and state shape functions are chosen so as to verify an homogeneous Dirichlet condition. 
\end{remark}

\subsection{Application to thin planar beams}
\label{sec:ph_floatbeam}

\begin{figure}[t]
	\centering
	\includegraphics[width=0.5\textwidth]{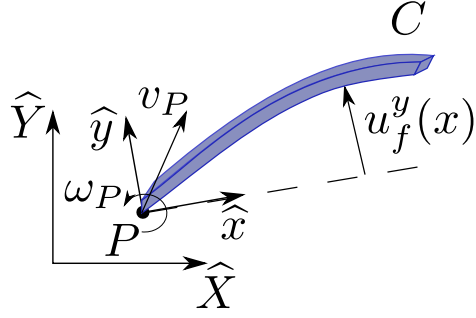} 
	\caption{Floating beam. The rigid motion is located at point P}
	\label{fig:beam}
\end{figure}

A thin planar flexible beam is considered as mechanical model. The dependence of the canonical momenta on the deformation field is neglected. This hypothesis usually applies in the floating frame formulation, since the deformations are small. $P$ is placed at the origin of the local frame $P=\{x=0\}$, while $C$ is the ending point of the beam $C=\{x=L\}$ (see Fig. \ref{fig:beam}). The beam has length $L$, Young modulus $E$, density $\rho$, cross section $A$ and second moment of area $I$. The model in strong form for a flexible beam is then written compactly as 
\begin{equation}
\label{eq:EB_str_phdae}
\begin{aligned}
\begin{bmatrix}
\bm{I} & 0 \\
0 & \bm{\mathcal{M}} \\
\end{bmatrix}
\begin{bmatrix}
\dot{\bm{q}} \\
\dot{\bm{p}} \\
\end{bmatrix}
&= \begin{bmatrix}
0 & \bm{\mathcal{J}}_{qe} \\
-\bm{\mathcal{J}}_{qe}^* & \bm{\mathcal{J}}_e \\
\end{bmatrix}
\begin{bmatrix}
\partial_{\bm{q}} H \\
\bm{p}  \\
\end{bmatrix} + 
\begin{bmatrix}
0 \\
\bm{\mathcal{B}}_r \\
\end{bmatrix} \bm{u}_\partial, \\
\bm{u}_\partial &= \bm{\mathcal{B}}_{\partial} \bm{p}, \\
\bm{y}_\partial &= \bm{\mathcal{C}}_{\partial} \bm{p},
\end{aligned}
\end{equation}
The state and boundary vectors are expressed as
\begin{align*}
\bm{q} &= [^i\bm{r}_P, \; \bm{R}_{\text{v}}, \; \bm{u}_f]^\top \\
\bm{p} &= [v_P^x, \; v_P^y, \; \omega_P^z, \; v_f^x, \; v_f^y, \; n_x, \; m_{x}]^\top, \\
\bm{u}_\partial &=  [F_{P}^x, \; F_{P}^y, \; T_{P}^z, \; F_{C}^x, \; F_{C}^y, \; T_{C}^z]^\top, \\
\bm{y}_\partial &=  [v_{P}^x, \; v_{P}^y, \; \omega_{P}^z, \; v_{C}^x, \; v_{C}^y, \; \omega_{C}^z]^\top.
\end{align*}
The state contains the generalized coordinates $\bm{q}$, the linear and angular velocity $v_P^x, \; v_P^y, \; \omega_P^z$ at point $P$, the deformation velocity $ v_f^x, \; v_f^y$ and the traction and bending stress $n_x, \; m_{x}$. The boundary input contains the forces and torques acting at the extremities of the beam, while the boundary output contains the corresponding conjugated variables (velocities and angular velocities).
The deformation field has to be constrained, to prevent rigid movement (see Rmk. \ref{rmk:dirich}). The appropriate selection of the boundary condition for the deformation field is an unavoidable problem that depends on the particular case under consideration.  Depending on the application, cantilever or simply supported boundary conditions may be considered (see Sec. \secref{sec:valid})
\begin{equation*}
\text{Cantilever}
\begin{cases}
u_f^x(x=0) = 0, \\
u_f^y(x=0) = 0, \\
\partial_x u_f^y(x=0) = 0, \\
\end{cases} \qquad 
\text{Simply supported}
\begin{cases}
u_f^x(x=0) = 0, \\
u_f^y(x=0) = 0, \\
u_f^y(x=L) = 0. \\
\end{cases}
\end{equation*}
Partitioning the $\bm{p}$ vector into rigid $\bm{p}_r = [v_P^x, \ v_P^y, \ \omega_P^z]^\top$ and flexible part $\bm{p}_f = [v_f^x, \ v_f^y, \ n_x, \ m_{x}]^\top$, the mass operator is then formulated as follows
\begin{equation}
\label{eq:EB_M}
\bm{\mathcal{M}} = 
\left[ \begin{array}{c:c}
\bm{\mathcal{M}}_{rr} & \bm{\mathcal{M}}_{rf} \\
\hdashline
\bm{\mathcal{M}}_{fr} & \bm{\mathcal{M}}_{ff} \\
\end{array} \right] = 
\left[ \begin{array}{ccc:cccc}
m & 0 & 0 & \mathcal{I}_\rho^L & 0 & 0 & 0 \\
0 & m & s^x & 0 & \mathcal{I}_\rho^L & 0 & 0 \\
0 & s^x & J^{zz} & 0 & \mathcal{I}_{\rho x}^{L} & 0 & 0 \\
\hdashline 
(\mathcal{I}_\rho^{L})^* & 0 & 0 & \rho A & 0 & 0 & 0  \\
0 & (\mathcal{I}_\rho^{L})^* & (\mathcal{I}_{\rho x}^{L})^* & 0 & \rho A & 0 & 0  \\
0 & 0 & 0 & 0 & 0 & {EA}^{-1} & 0 \\
0 & 0 & 0 & 0 & 0 & 0 & {EI}^{-1} \\
\end{array} \right],
\end{equation}
where  $s^x = \int_{0}^{L} \rho A x \d{x}$ is the static moment, $J^{zz} = \int_{0}^{L} \rho A x^2 \d{x}$ is the moment of inertia, $\mathcal{I}_\rho^L := \int_{0}^L \rho A (\cdot) \d{x}, \; \mathcal{I}_{\rho x}^L := \int_{0}^L \rho A x (\cdot) \d{x}$. The interconnection operator is found by adapting the cross product to the planar case:
\begin{equation}
\label{eq:EB_J}
\bm{\mathcal{J}}_{e}(\bm{e}) = 
\left[ \begin{array}{c:c}
\bm{\mathcal{J}}_{rr} & \bm{\mathcal{J}}_{rf} \\
\hdashline
\bm{\mathcal{J}}_{fr} & \bm{\mathcal{J}}_{ff} \\
\end{array} \right] = 
\left[ \begin{array}{ccc:cccc}
0 & 0 & +\widetilde{p}_t^y      & 0 & 0 & 0 & 0 \\
0 & 0 & -\widetilde{p}_t^x     & 0 & 0 & 0 & 0 \\
-\widetilde{p}_t^y & +\widetilde{p}_t^x & 0 & -\mathcal{I}_{p_f^y}^{L} & +\mathcal{I}_{p_f^x}^{L} & 0 & 0 \\
\hdashline 
0 & 0 & +(\mathcal{I}_{p_f^y}^{L})^* & 0 & 0 & \partial_x & 0  \\
0 & 0 & -(\mathcal{I}_{p_f^x}^{L})^* & 0 & 0 & 0 & -\partial_{xx} \\
0 & 0 & 0 & \partial_{x} & 0 & 0 & 0 \\
0 & 0 & 0 & 0 & \partial_{xx} & 0 & 0 \\
\end{array} \right],
\end{equation}
where $\widetilde{p}_t^x, \widetilde{p}_t^y$ are the modified canonical momenta components (see \eqref{eq:mod_momenta}), $\mathcal{I}_{p_f^x}^{L} := \int_{0}^{L} \left\{2 p_f^x + \rho A v_f^x \right\} (\cdot) \d{x}$ and $\mathcal{I}_{p_f^y}^{L} := \int_{0}^{L} \left\{2 p_f^y + \rho A v_f^y \right\} (\cdot) \d{x}$. The control operator reads
\begin{equation}
\bm{\mathcal{B}}_r = \begin{bmatrix}
\bm{I}_{3\times 3} & \bm\tau_{CP}^\top \\
0_{4\times 3} & 0_{4\times 3} \\
\end{bmatrix} \qquad \text{with} \quad
\bm\tau_{CP} = \begin{bmatrix}
1 & 0 & 0 \\
0 & 1 & L \\
0 & 0 & 1 \\
\end{bmatrix}.
\end{equation}

The discretization procedure detailed in \secref{sec:discr} is extended to this case, considering that the differential operators are 
\[
\bm{\mathcal{J}}_{\Div} = \begin{bmatrix}
0 & 0 & \partial_x & 0 \\
0 & 0 & 0 & -\partial_{xx} \\
0 & 0 & 0 & 0 \\
0 & 0 & 0 & 0 \\
\end{bmatrix}, \qquad 
\bm{\mathcal{J}}_{\Grad} = \begin{bmatrix}
0 & 0 & 0 & 0 \\
0 & 0 & 0 & 0 \\
\partial_x & 0 & 0 & 0 \\
0 & \partial_{xx} & 0 & 0 \\
\end{bmatrix}.
\]
These two operators play the same role as their previously defined homonyms. The 2 PDEs associated to the first and second line of $\bm{\mathcal{J}}_{\Div}$ are integrated by parts once and twice respectively, so that the boundary forces and momenta are naturally included in the discretized system as inputs. The finite-dimensional system then reads
\begin{equation}
\label{eq:EB_sys}
\begin{aligned}
\begin{bmatrix}
\mathbf{I} & 0 & 0 \\
0 & \mathbf{M}_{rr} & \mathbf{M}_{rf}\\ 
0 & \mathbf{M}_{fr} & \mathbf{M}_{ff}\\
\end{bmatrix} 
\begin{bmatrix}
\dot{\mathbf{q}} \\ 
\dot{\mathbf{p}}_{r} \\ 
\dot{\mathbf{p}}_{f} \\ 
\end{bmatrix} &= 
\begin{bmatrix}
0 & \mathbf{J}_{qr}(\mathbf{q}) & \mathbf{J}_{qf} \\
\mathbf{J}_{rq}(\mathbf{q}) & \mathbf{J}_{rr}(\mathbf{p}) & \mathbf{J}_{rf}(\mathbf{p})\\ 
\mathbf{J}_{fq} & \mathbf{J}_{fr}(\mathbf{p}) & \mathbf{J}_{ff}\\
\end{bmatrix}  
\begin{bmatrix}
\partial_{\mathbf{q}} H \\
{\mathbf{p}}_{r} \\ 
{\mathbf{p}}_{f} \\ 
\end{bmatrix} + 
\begin{bmatrix}
0 \\
\mathbf{B}_{r} \\ 
\mathbf{B}_{f} \\ 
\end{bmatrix}
\mathbf{u}_\partial, \\
\mathbf{y}_\partial &= \begin{bmatrix}
0 \ & \mathbf{B}_{r}^\top & \mathbf{B}_{f}^\top  
\end{bmatrix} \begin{bmatrix}
\mathbf{q} \\
{\mathbf{p}}_{r} \\ 
{\mathbf{p}}_{f} \\ 
\end{bmatrix},
\end{aligned}
\end{equation}
Matrix $\mathbf{B}_{r} = [\mathbf{I}_{3\times 3}, \; \bm\tau_{CP}^\top]$ accounts for the effect of boundary forces on the rigid part. Matrix $\mathbf{B}_{f}$  is the result of the integration by parts
\begin{align*}
\mathbf{B}_{f} = \begin{bmatrix}
0_{n_f^{vx} \times 3} & \bm{\phi}_{v_f^x}(L) & 0_{n_f^{vx}} & 0_{n_f^{vx}} \\
0_{n_f^{vy} \times 3} & 0_{n_f^{vy}} & \bm{\phi}_{v_f^y}(L) & \partial_x \bm{\phi}_{v_f^y}(L) \\
0_{n_f^{\sigma x} \times 3} & 0_{n_f^{\sigma x}} & 0_{n_f^{\sigma x}} & 0_{n_f^{\sigma x}} \\
0_{n_f^{\sigma y} \times 3} & 0_{n_f^{\sigma y}} & 0_{n_f^{\sigma y}} & 0_{n_f^{\sigma y}} \\
\end{bmatrix},
\end{align*}
where $\bm{\phi}_{v_f^x}, \ \bm{\phi}_{v_f^y}$ are the shape functions for ${v}_f^x, {v}_f^y$ and $\bm{\phi}_{v_f^x}$. Fields ${v}_f^x, {v}_f^y, n_x, m_x$ are approximated using $n_f^{vx}, n_f^{vy}, n_f^{\sigma x}, n_f^{\sigma y}$ degrees of freedom respectively. System \eqref{eq:EB_sys} can be rewritten compactly as
\begin{equation}
\label{eq:beam_discr}
\begin{aligned}
\mathbf{E} \dot{\mathbf{e}} &= \mathbf{J}(\mathbf{e}) \mathbf{z}(\mathbf{e}) + \mathbf{B}_\partial \mathbf{u}_\partial, \vspace{2mm} \\
\mathbf{y}_\partial &= \mathbf{B}_\partial^{\top}  \mathbf{z}. \\
\end{aligned}
\end{equation}
This model describes the motion of a flexible floating beam that undergoes small deformations. 

\section{Multibody systems in pH form}
\label{sec:MB_pH}
In Sections \secref{sec:pH_fd}, and \secref{sec:discr}, the pH formulation of a single flexible floating body in infinite- and finite-dimensional form was presented. The construction of a multibody system is accomplished by exploiting the modularity of the port-Hamiltonian framework. Each element of the system is interconnected to the others by means of classical pH interconnections.

\subsection{Interconnections of pHDAE systems}
Consider two generic pHDAE systems of the form
\begin{equation}
\begin{cases}
\mathbf{E}_i \dot{\mathbf{e}}_i = \mathbf{J}_i \mathbf{z}_i(\mathbf{e}_i) + \mathbf{B}_i^{\text{int}} \mathbf{u}_i^{\text{int}} + \mathbf{B}_i^{\text{ext}} \mathbf{u}_i^{\text{ext}}  \vspace{2mm} \\
\mathbf{y}_i^{\text{int}} = \mathbf{B}_i^{\text{int} \top}  \mathbf{z}_i \\
\mathbf{y}_i^{\text{ext}} = \mathbf{B}_i^{\text{ext} \top}  \mathbf{z}_i \\
\end{cases} \qquad \forall i = 1, 2.
\end{equation}
where $\partial_{\mathbf{e}_i} {H_i} = \mathbf{E}_i^\top \mathbf{z}_i$. Systems of this kind arise from the discretization of formulation \eqref{eq:MFD_pHDAE}. The interconnection uses the internal control $\mathbf{u}_i^{\text{int}}$. An interconnection is said to be power preserving if and only if the following holds
\begin{equation} \label{eq:int_balance}
\langle \mathbf{u}_1^{\text{int}}, \; \mathbf{y}_1^{\text{int}} \rangle + \langle \mathbf{u}_2^{\text{int}}, \; \mathbf{y}_2^{\text{int}} \rangle = 0,
\end{equation}

which expresses that the power going out from one system flows in the other in a lossless manner. Two interconnections are of interest when coupling system: the gyrator and transformer interconnections.

\paragraph{Gyrator interconnection}
The gyrator interconnection reads
\begin{equation*}
\mathbf{u}_1^{\text{int}} = -\mathbf{C} \mathbf{y}_2^{\text{int}}, \qquad
\mathbf{u}_2^{\text{int}} = \mathbf{C}^\top \mathbf{y}_1^{\text{int}}.
\end{equation*}
This interconnection verifies \eqref{eq:int_balance} and provides the system
\begin{align*}
\begin{bmatrix}
\mathbf{E}_1 & 0 \\ 0 & \mathbf{E}_2 \\
\end{bmatrix}
\begin{bmatrix}
\dot{\mathbf{e}}_1 \\ \dot{\mathbf{e}}_2 \\
\end{bmatrix} &= 
\begin{bmatrix}
\mathbf{J}_1 & -\mathbf{B}_1^{\text{int}} \mathbf{C} \mathbf{B}_2^{\text{int} \top} \\ 
\mathbf{B}_2^{\text{int}} \mathbf{C} \mathbf{B}_1^{\text{int} \top}  & \mathbf{J}_2 \\
\end{bmatrix}
\begin{bmatrix}
\mathbf{z}_1 \\ 
\mathbf{z}_2 \\
\end{bmatrix}+ 
\begin{bmatrix}
\mathbf{B}_1^{\text{ext}} & 0 \\ 0 & \mathbf{B}_2^{\text{ext}} \\
\end{bmatrix} 
\begin{bmatrix}
\mathbf{u}_1^{\text{ext}} \\ \mathbf{u}_2^{\text{ext}} \\
\end{bmatrix}  \\
\begin{bmatrix}
\mathbf{y}_1^{\text{ext}} \\ \mathbf{y}_2^{\text{ext}} \\
\end{bmatrix}  &= \begin{bmatrix}
\mathbf{B}_1^{\text{ext} \top} & 0 \\
0 & \mathbf{B}_2^{\text{ext} \top} \\
\end{bmatrix} \begin{bmatrix}
\mathbf{z}_1 \\ 
\mathbf{z}_2 \\
\end{bmatrix}.
\end{align*}

\paragraph{Transformer interconnection}
The transformer interconnection reads
\begin{equation*}
\mathbf{u}_1^{\text{int}} = -\mathbf{C} \mathbf{u}_2^{\text{int}}, \qquad
\mathbf{y}_2^{\text{int}} = \mathbf{C}^\top \mathbf{y}_1^{\text{int}}.
\end{equation*}
Again, this interconnection verifies \eqref{eq:int_balance}. After the interconnection the final system is differential algebraic:
\begin{align*}
\begin{bmatrix}
\mathbf{E}_1 & 0 & 0 \\ 
0 & \mathbf{E}_2 & 0 \\
0 & 0 & 0 \\
\end{bmatrix}
\begin{bmatrix}
\dot{\mathbf{e}}_1 \\ \dot{\mathbf{e}}_2 \\ \dot{\bm{\lambda}} \\
\end{bmatrix} &= 
\begin{bmatrix}
\mathbf{J}_1 & 0 & -\mathbf{B}_1^{\text{int}} \mathbf{C} \\ 
0 & \mathbf{J}_2 & \mathbf{B}_2^{\text{int}} \\
\mathbf{C}^\top \mathbf{B}_1^{\text{int} \top} & - \mathbf{B}_2^{\text{int} \top} & 0 \\
\end{bmatrix}
\begin{bmatrix}
\mathbf{z}_1 \\ 
\mathbf{z}_2 \\
\bm{\lambda} \\
\end{bmatrix}+ 
\begin{bmatrix}
\mathbf{B}_1^{\text{ext}} & 0 \\ 0 & \mathbf{B}_2^{\text{ext}} \\ 0 & 0 \\
\end{bmatrix} 
\begin{bmatrix}
\mathbf{u}_1^{\text{ext}} \\ 
\mathbf{u}_2^{\text{ext}} \\
\end{bmatrix} \\
\begin{bmatrix}
\mathbf{y}_1^{\text{ext}} \\ \mathbf{y}_2^{\text{ext}} \\
\end{bmatrix}  &= \begin{bmatrix}
\mathbf{B}_1^{\text{ext} \top} & 0 & 0 \\
0 & \mathbf{B}_2^{\text{ext} \top} & 0 \\
\end{bmatrix} \begin{bmatrix}
\mathbf{z}_1 \\ 
\mathbf{z}_2 \\
\bm{\lambda} \\
\end{bmatrix}.
\end{align*}

\subsection{Application to multibody systems of beams}
\label{sec:int_beams}
Once a discretized system is obtained, lossless joints can be modeled as a transformer interconnection. A common example is a revolute joint between two beams. Considering discretization \eqref{eq:beam_discr}, the boundary control input $\mathbf{u}_{\partial, i}$ may be split into interconnection variables $\mathbf{u}_i^{\text{int}}$ and external variables  $\mathbf{u}_i^{\text{ext}}$, i.e. $\mathbf{u}_{\partial, i} = [\mathbf{u}_i^{\text{int}}; \ \mathbf{u}_i^{\text{ext}}]$. The same splitting applies to the output. In this case the interconnection variables are
\begin{equation*}
\begin{aligned}
\mathbf{u}_1^{\text{int}} &= [F^x_{C_1}, \, F^y_{C_1}]^\top := \mathbf{F}_{C_1}, \\
\mathbf{u}_2^{\text{int}} &= [F^x_{P_2}, \, F^y_{P_2}]^\top := \mathbf{F}_{P_2},
\end{aligned} \qquad
\begin{aligned}
\mathbf{y}_1^{\text{int}} &= [v^x_{C_1}, \, v^y_{C_1}]^\top := \mathbf{v}_{C_1}, \\
\mathbf{y}_2^{\text{int}} &= [v^x_{P_2}, \, v^y_{P_2}]^\top := \mathbf{v}_{P_2}.
\end{aligned}
\end{equation*}
The interconnection matrix is the relative rotation matrix between the two local frames
\begin{equation}
\mathbf{R}(\theta) = \begin{bmatrix}
\cos(\theta) & - \sin(\theta) \\
\sin(\theta) & \cos(\theta) \\
\end{bmatrix}, \qquad 
\begin{aligned}
\theta(t) &= \theta(0) + \int_{0}^t (\omega^z_{P_2} - \omega^z_{P_1}) \d{\tau}.
\end{aligned}
\end{equation}

\begin{figure}[t]
	\centering
	\includegraphics[width=0.45\textwidth]{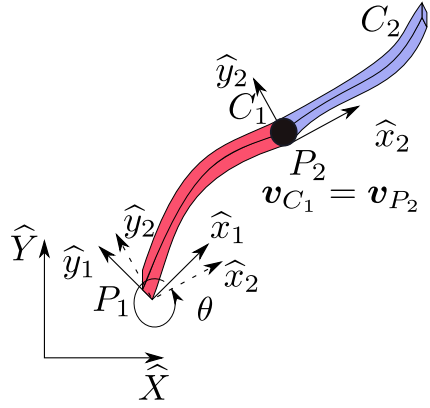} 
	
	\caption{Two beams interconnected by an hinge }
	\label{fig:beam_int}
\end{figure}

The transformer interconnection
\begin{equation}
\label{eq:int_hinge}
\mathbf{u}_1^{\text{int}} = -\mathbf{R}(\theta) \mathbf{u}_2^{\text{int}}, \qquad
\mathbf{y}_2^{\text{int}} = \mathbf{R}(\theta)^\top \mathbf{y}_1^{\text{int}},
\end{equation}
imposes the constraints on the velocity level and gives rise to a quasi-linear index 2 pHDAE (see Appendix A for the index definition):

\begin{equation}
\label{eq:int_beams}
\begin{aligned}
\begin{bmatrix}
\mathbf{E}_1 & 0 & 0 \\ 
0 & \mathbf{E}_2 & 0 \\
0 & 0 & 0 \\
\end{bmatrix}
\begin{bmatrix}
\dot{\mathbf{e}}_1 \\ \dot{\mathbf{e}}_2 \\ \dot{\bm{\lambda}} \\
\end{bmatrix} &= 
\begin{bmatrix}
\mathbf{J}_1(\mathbf{e}_1) & 0 & -\mathbf{B}_1^{\text{int}} \mathbf{R} \\ 
0 & \mathbf{J}_2(\mathbf{e}_2) & \mathbf{B}_2^{\text{int}} \\
\mathbf{R}^\top \mathbf{B}_1^{\text{int} \top} & - \mathbf{B}_2^{\text{int} \top} & 0 \\
\end{bmatrix}
\begin{bmatrix}
\mathbf{z}_1  \\ 
\mathbf{z}_2  \\ 
\bm{\lambda} \\
\end{bmatrix}+ 
\begin{bmatrix}
\mathbf{B}_{\partial 1}^{\text{ext}} & 0 \\ 0 & \mathbf{B}_{\partial 2}^{\text{ext}} \\ 0 & 0 \\
\end{bmatrix} 
\begin{bmatrix}
\mathbf{u}_1^{\text{ext}} \\ 
\mathbf{u}_2^{\text{ext}} \\
\end{bmatrix}, \\
\begin{bmatrix}
\mathbf{y}_1^{\text{ext}} \\ \mathbf{y}_2^{\text{ext}} \\
\end{bmatrix}  &= \begin{bmatrix}
\mathbf{B}_{\partial 1}^{\text{ext} \top} & 0 & 0 \\
0 & \mathbf{B}_{\partial 2}^{\text{ext} \top} & 0 \\
\end{bmatrix} \begin{bmatrix}
\mathbf{z}_1  \\ 
\mathbf{z}_2  \\ 
\bm{\lambda} \\
\end{bmatrix}.
\end{aligned}
\end{equation}

\begin{figure}[t]
	\centering
	\includegraphics[height=0.16\textheight]{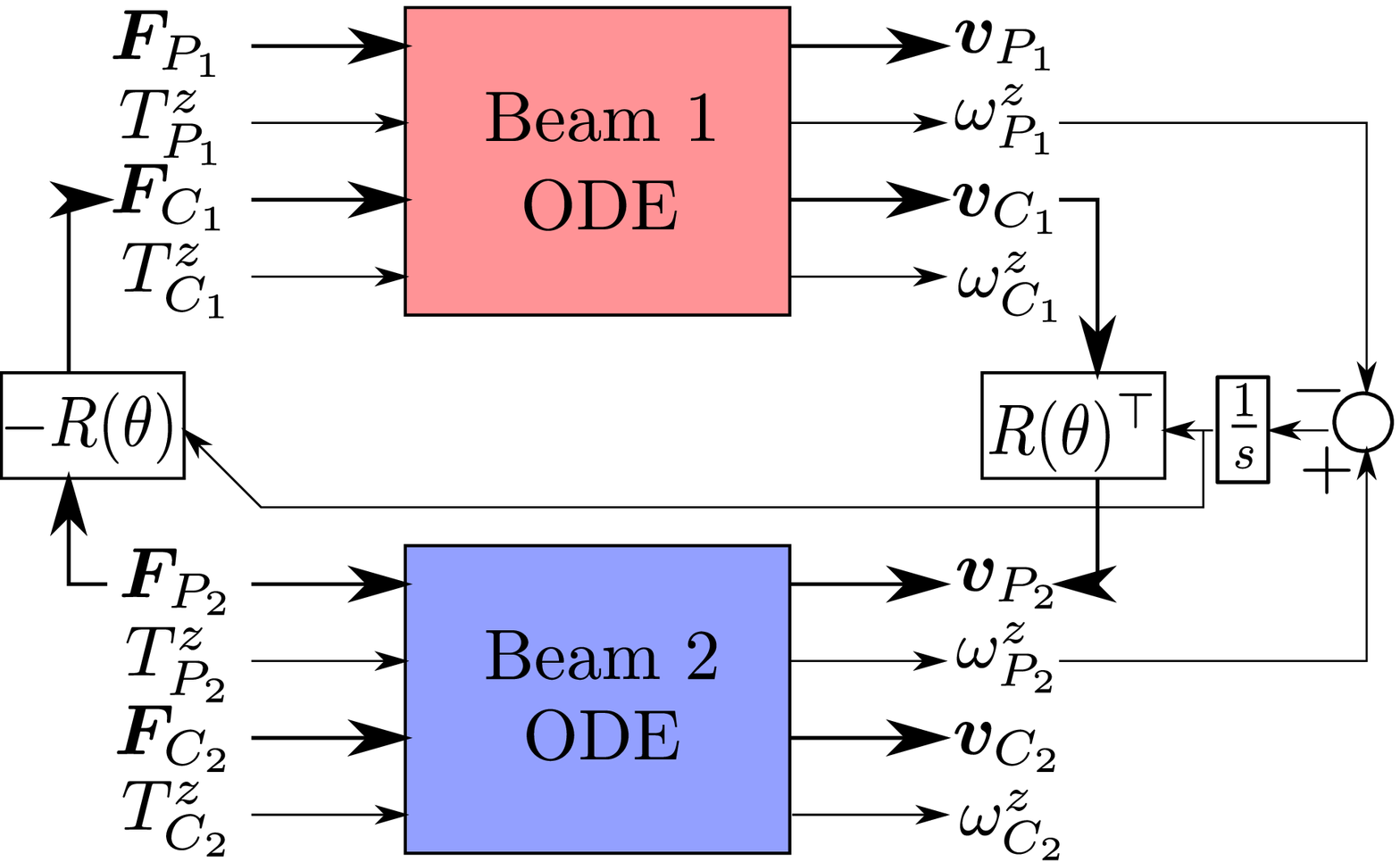} \hspace{.5cm}
	\includegraphics[height=0.16\textheight]{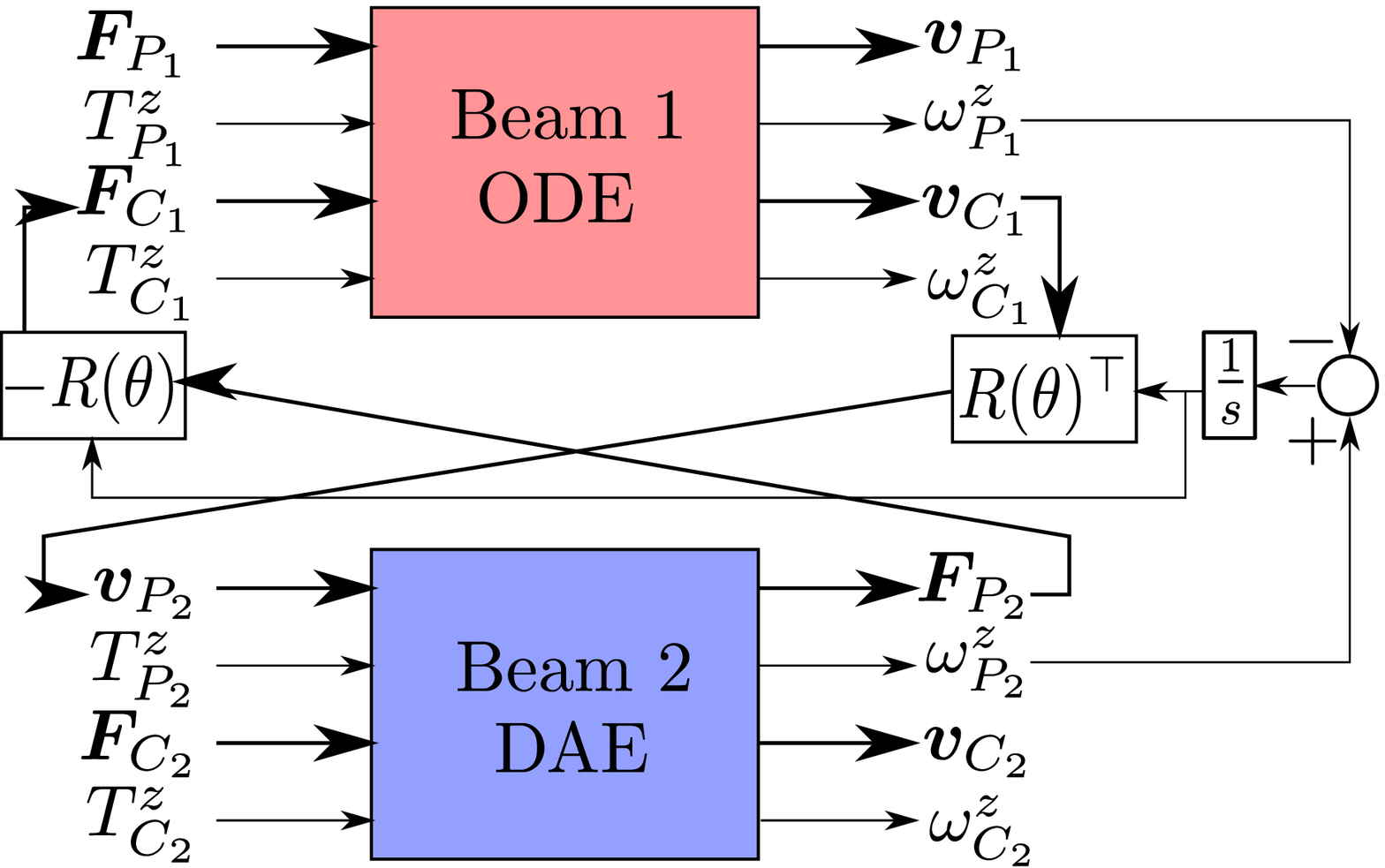} 	
	\caption{Block diagrams representing the transformer interconnection \eqref{eq:int_hinge} (left) and the equivalent gyrator interconnection \eqref{eq:int_hinge_DAE} (right)}
	\label{fig:beam_int_block}
\end{figure}

The same result can be obtained by using a pHDAE system and a gyrator interconnection. To illustrate this, consider the pHDAE obtained by interchanging the role of output and input of the second system $\mathbf{u}_2^{\text{int}} \leftrightarrow \mathbf{y}_2^{\text{int}}$. The output then plays the role of a Lagrange multiplier. The input $\mathbf{u}_2^{\text{int}}$ is now considered as Lagrange multiplier ${\bm{\lambda}}_2$ and the output $\mathbf{y}_2^{\text{int}}$ plays the role of $\mathbf{u}_2^{\text{int}}$. The discretized system assumes the following differential-algebraic structure
\begin{equation}
\begin{aligned}
\begin{bmatrix}
\mathbf{E}_2 & 0 \\
0 & 0 \\
\end{bmatrix} \begin{bmatrix}
\dot{\mathbf{e}}_2 \\
\dot{\bm{\lambda}}_2 \\
\end{bmatrix}
&= \begin{bmatrix}
\mathbf{J}_2(\mathbf{e}_2) & \mathbf{B}_2^{\text{int}} \\
-\mathbf{B}_2^{\text{int} \top} & 0\\
\end{bmatrix} \begin{bmatrix}
\mathbf{z}_2 \\
{\bm{\lambda}}_2
\end{bmatrix} 
+ \begin{bmatrix}
0 \\
\mathbf{I}
\end{bmatrix} \mathbf{u}_2^{\text{int}} + \begin{bmatrix}
\mathbf{B}_2^{\text{ext}} \\
0 \\
\end{bmatrix} \mathbf{u}_2^{\text{ext}},  \vspace{2mm} \\
\mathbf{y}_2^{\text{int}} &= {\bm{\lambda}}_2, \\
\mathbf{y}_2^{\text{ext}} &= \mathbf{B}_2^{\text{ext} \top} \mathbf{z}_2.
\end{aligned}
\end{equation}
This system is improper, since the input appears in the algebraic part. Now, a gyrator interconnection is used to model the hinged joint
\begin{equation}
\label{eq:int_hinge_DAE}
\mathbf{u}_1^{\text{int}} = -\mathbf{R}(\theta) \mathbf{y}_2^{\text{int}}, \qquad
\mathbf{u}_2^{\text{int}} = \mathbf{R}(\theta)^\top \mathbf{y}_1^{\text{int}}.
\end{equation}
The resulting differential-algebraic system is exactly \eqref{eq:int_beams}, which is proper. The equivalence between the two representation is represented in Fig. \ref{fig:beam_int_block}. This approach allows the modular construction of systems of arbitrary complexity. Other kind of lossless joints (prismatic, spherical) can be modeled by appropriate interconnections. The system can then be simulated by using specific DAE solvers \cite{daePetzold}. 

\subsection{The linear case: sub-structuring and model reduction}
If the angular velocities and the relative orientations are small then the system may be linearized about a particular geometrical configuration. Omitting the partition related to the generalized coordinates $\mathbf{q}$ and partitioning the system into rigid and flexible dynamics, the resulting equations are then expressed as 
\begin{equation}
\label{eq:mbd_linear}
\begin{bmatrix}
\mathbf{M}_{rr} & \mathbf{M}_{rf} & 0 \\ 
\mathbf{M}_{fr} & \mathbf{M}_{ff} & 0 \\
0 & 0 & 0 \\
\end{bmatrix}
\begin{bmatrix}
\dot{\mathbf{p}}_r \\ \dot{\mathbf{p}}_f \\ \dot{\bm{\lambda}} \\ 
\end{bmatrix} = 
\begin{bmatrix}
0 & 0 & \mathbf{G}_r^\top \\ 
0 & \mathbf{J}_{ff} & \mathbf{G}_f^\top \\ 
-\mathbf{G}_r & -\mathbf{G}_f & 0 \\
\end{bmatrix}
\begin{bmatrix}
\mathbf{p}_r \\ \mathbf{p}_f \\ {\bm{\lambda}} \\ 
\end{bmatrix} + 
\begin{bmatrix}
\mathbf{B}_r \\ \mathbf{B}_f \\ 0 \\
\end{bmatrix}\mathbf{u}.
\end{equation}
The Hamiltonian is now a quadratic function of the state variables $H = \frac{1}{2} \mathbf{p}^\top\mathbf{M}\mathbf{p}$ \cite{beattie2018linear}.
The modular construction of complex multi-body systems is then analogous to a sub-structuring technique \cite{substructuring}, where the velocities and forces are linked at the interconnection points. Such system can be reduced using Krylov subspace method directly on the DAE formulation \cite{phdae_red}. The basic idea relies on the construction of a subspace $\mathbf{V}_f^{\text{red}}$ for the vector $\mathbf{p}_f$ such that $\mathbf{p}_f \approx \mathbf{V}_f^{\text{red}} \mathbf{p}_f^{\text{red}}$. The reduced system then reads
\begin{equation}
\begin{bmatrix}
\mathbf{M}_{rr} & \mathbf{M}_{rf}^{\text{red}} & 0 \\ 
\mathbf{M}_{fr}^{\text{red}} & \mathbf{M}_{ff}^{\text{red}} & 0 \\
0 & 0 & 0 \\
\end{bmatrix}
\begin{bmatrix}
\dot{\mathbf{p}}_r \\ \dot{\mathbf{p}}_f^{\text{red}} \\ \dot{\mathbf{\lambda}} \\ 
\end{bmatrix} = 
\begin{bmatrix}
0 & 0 & \mathbf{G}_r^\top \\ 
0 & \mathbf{J}_{ff}^{\text{red}} & \mathbf{G}_f^{\text{red} \top} \\ 
-\mathbf{G}_r & -\mathbf{G}_f^{\text{red}} & 0 \\
\end{bmatrix}
\begin{bmatrix}
\mathbf{p}_r \\ \mathbf{p}_f^{\text{red}} \\ {\mathbf{\lambda}} \\ 
\end{bmatrix} + 
\begin{bmatrix}
\mathbf{B}_r \\ \mathbf{B}_f^{\text{red}} \\ 0 \\
\end{bmatrix}\mathbf{u},
\end{equation}
where the second row has been pre-multiplied by $\mathbf{V}_f^{\text{red} \top}$. Alternatively, a null space matrix can employed to eliminate the Lagrange multiplier and preserve the port-Hamiltonian structure. Consider the pHDAE \eqref{eq:mbd_linear}, where the differential and algebraic parts are explicitly separated

\begin{equation}
\begin{aligned}
\mathbf{M} \dot{\mathbf{p}} &=  \mathbf{J}\mathbf{p} + \mathbf{G}^\top \bm{\lambda} + \mathbf{B}\mathbf{u}, \\ 
0 &= \mathbf{G}\mathbf{p},
\end{aligned}
\end{equation}
and consider a matrix $\mathbf{P}$ that satisfies 
\[
\mathrm{range}\{\mathbf{P}\} = \mathrm{null}\{\mathbf{G}\}.
\]
Then, the range of $\mathbf{P}$ automatically satisfies the constraints. Considering the transformation $\widehat{\mathbf{p}} = \mathbf{P} \mathbf{p}$ and pre-multiplying the system by $\mathbf{P}^\top$ an equivalent ODE is obtained
\[
\widehat{\mathbf{M}} \ \dot{\widehat{\mathbf{p}}} =  \widehat{\mathbf{J}} \ \widehat{\mathbf{p}} + \widehat{\mathbf{B}} \ \mathbf{u},
\]
with $\widehat{\mathbf{M}} = \mathbf{P}^\top \mathbf{M} \mathbf{P}, \; \widehat{\mathbf{J}} = \mathbf{P}^\top \mathbf{J} \mathbf{P}, \; \widehat{\mathbf{B}} = \mathbf{P}^\top \mathbf{B}$. The computation of $\mathbf{P}$ can be performed by QR decomposition of matrix $\mathbf{G}$ \cite{nullspaceFlMult}. A pH system in standard form is then obtained considering the variable change $\widehat{\mathbf{x}} = \widehat{\mathbf{M}} \widehat{\mathbf{p}}$
\[ \dot{\widehat{\mathbf{x}}} =  \widehat{\mathbf{J}} \widehat{\mathbf{Q}}\ \widehat{\mathbf{x}} + \widehat{\mathbf{B}}  \mathbf{u}, \qquad \widehat{\mathbf{Q}}:= \widehat{\mathbf{M}}^{-1}.
\] 
Once an equivalent ODE formulation is obtained the concepts and ideas presented in \cite{phode_red} can be used to reduce the flexible dynamics.

\section{Validation}
\label{sec:valid}

In this section numerical simulations are performed to assess the correctness of the proposed formulation. A first example concerns the computation of eigenvalues of a four bar mechanics for different geometrical configuration. The second example is a rotating crank-slider. In this case the non-linearities cannot be neglected. The third example is a hinged beam undergoing external excitations so that the out-of-plane motion becomes important. The examples make use of Euler Bernoulli beam model \eqref{eq:EB_sys}. To discretize the system, Lagrange polynomial of order one are used for $v_f^x$ and $n_x$, while Hermite polynomials are used for $v_f^y$ and $m_{x}$. This choice ensures the conformity with respect to the differential operator. The Firedrake python library \cite{rathgeber2017firedrake} is employed to construct the finite-dimensional discretization.  

\subsection{Linear analysis of a four-bar mechanism}

\begin{figure}[tb]
	\centering
	\includegraphics[width=0.6\textwidth]{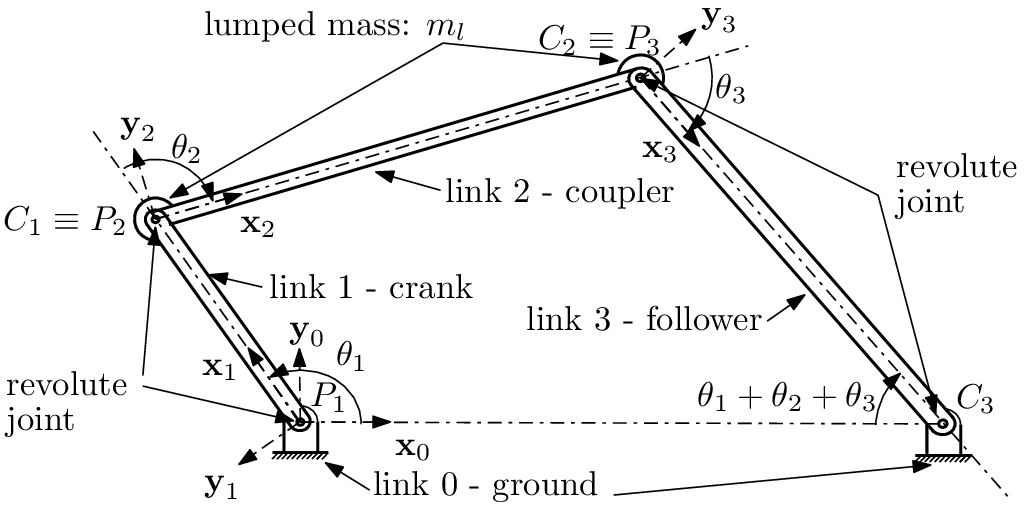} 
	\includegraphics[width=0.35\textwidth]{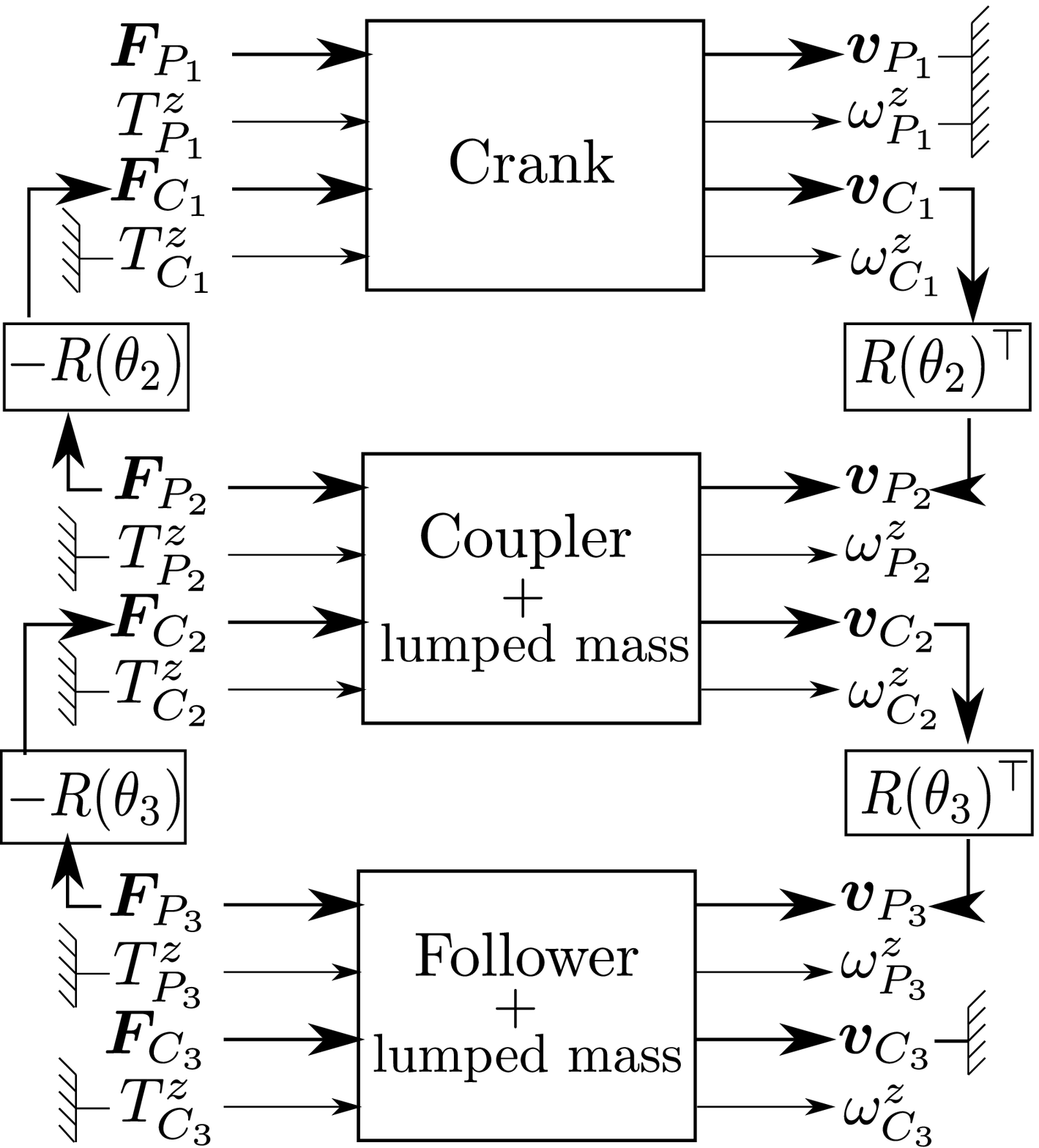} 
	\caption{Four bar mechanism illustration (left, taken from \cite{Chebbi2017}) and block diagram used for the eigenvalues analysis (right)}
	\label{fig:4bars}
\end{figure}

\begin{table}[bt]
	\caption{Four-bar mechanism links properties: each link is a uniform beam with mass density $\rho=2714\,[\mathrm{kg}/\mathrm{m}^3]$ and Young modulus $E=7.1\,10^{10}\,[\mathrm{N}/\mathrm{m}^2]$. The lumped masses $m_l=0.042\,[\mathrm{kg}]$ are taken into account considering an additional mass at $P$ for link 2 and 3.}
	\label{tab:data_4bars}       
	\begin{tabular}{lllll}
		\hline\noalign{\smallskip}
		$i$ & $0$ &  $1$ &  $2$ &  $3$  \\
		\noalign{\smallskip}\hline\noalign{\smallskip}
		Name & ground & crank & coupler & follower \\ 
		Length $L_i\,[\mathrm{m}]$ & $0.254$ & $0.108$ & $0.2794$ & $0.2705$\\
		Cross section $A_i\,[\mathrm{m}^2]$ & $-$ & $1.0774\,10^{-4}$ & $4.0645\,10^{-5}$ & $4.0645\,10^{-5}$ \\
		Flexural rigidity $(EI)_i\,[\mathrm{Nm}^2]$ & $-$ & $11.472$ & $0.616$ & $0.616$ \\
		\hline
	\end{tabular}
\end{table}

\begin{figure}[tb]
	\centering
	\includegraphics[width=0.49\textwidth]{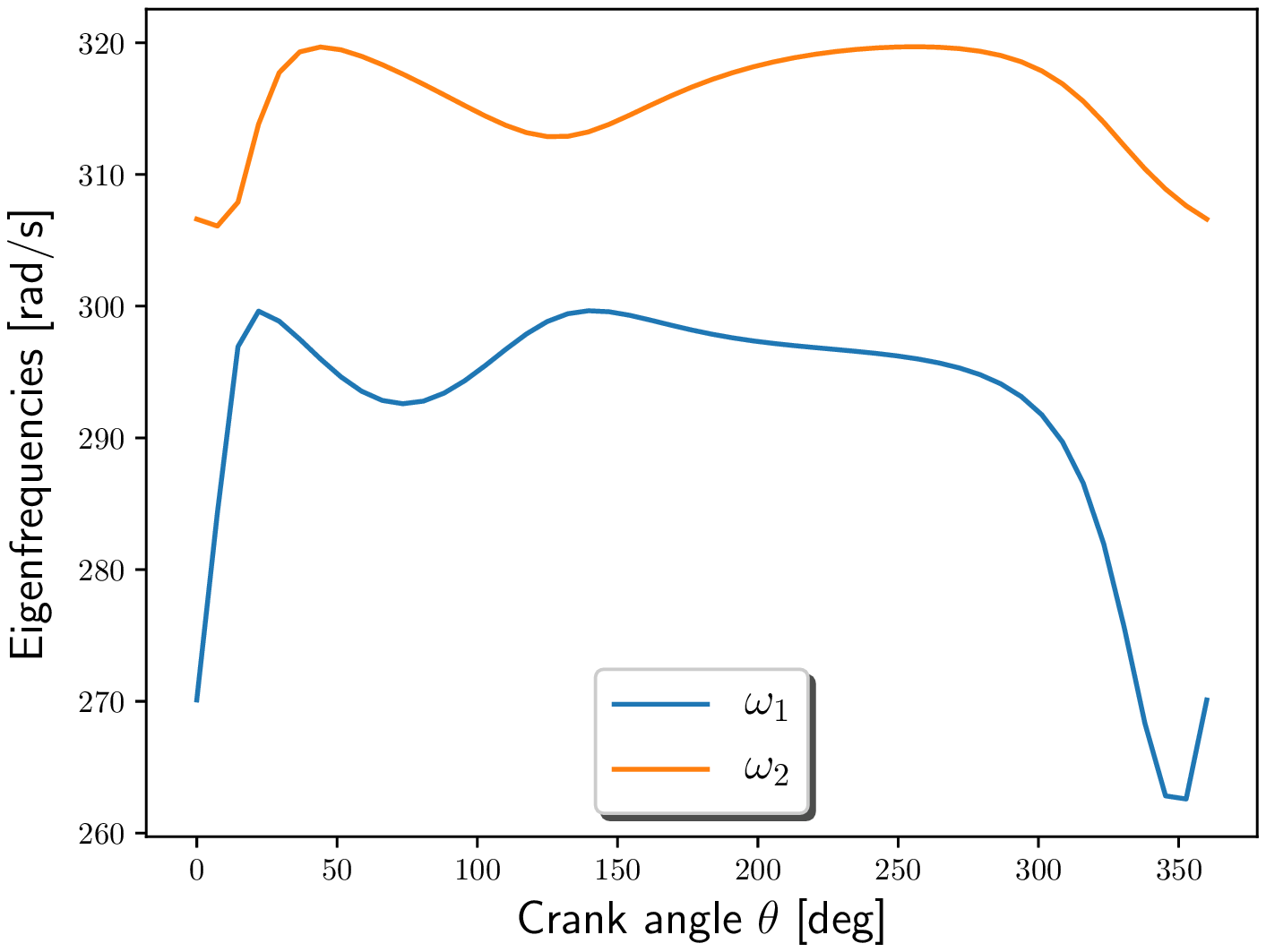} 
	\includegraphics[width=0.49\textwidth]{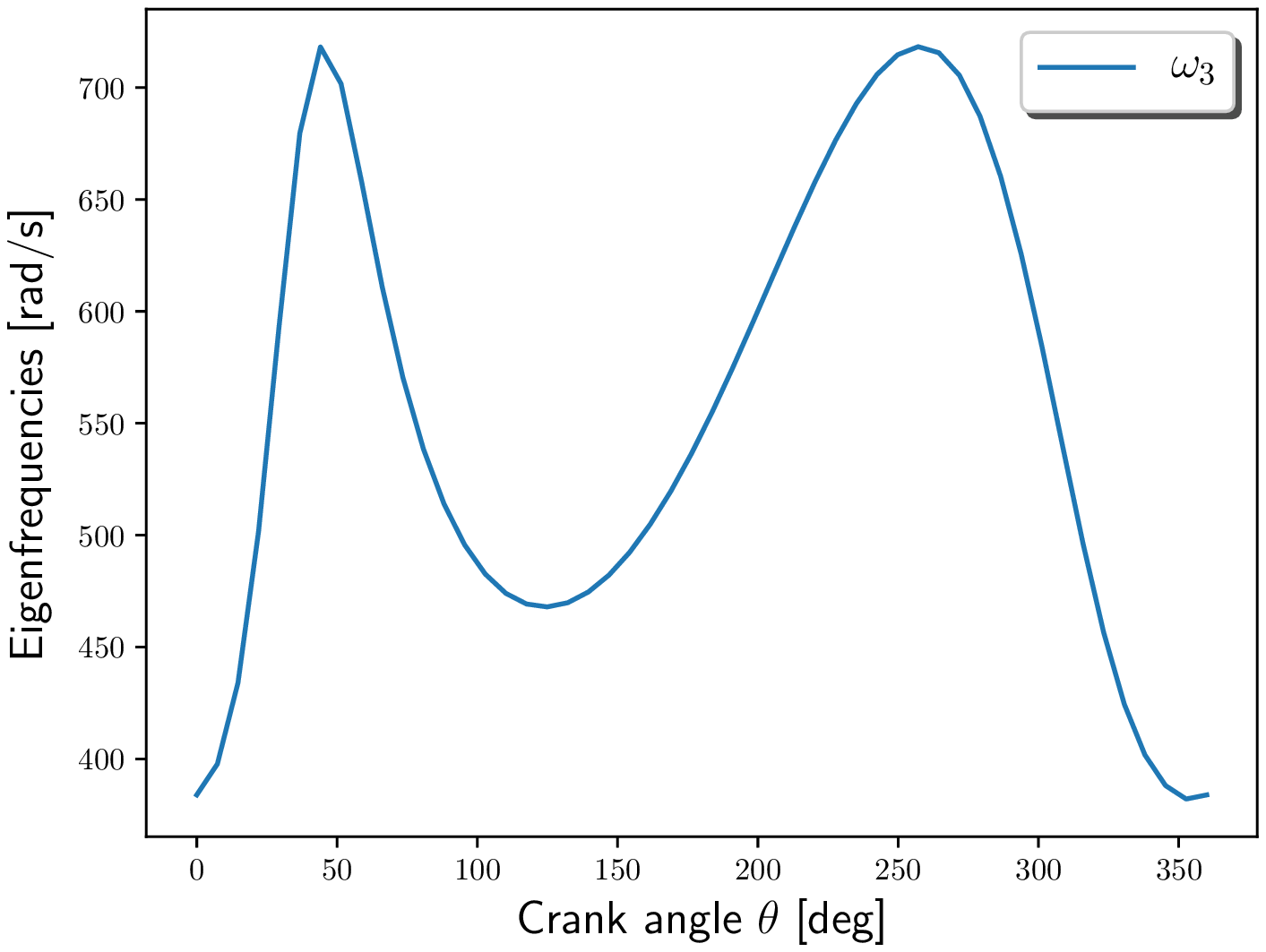} 
	\caption{Eigenvalues $\omega_i, \ i=1,2,3$ for the four bar mechanism for varying crank angle.}
	\label{fig:omega_4bars}
\end{figure}

The four-bar mechanism has one degree of freedom and represents a closed chain of bodies. The data are taken from \cite{KITIS1990267,Chebbi2017} are recalled in Table \ref{tab:data_4bars}. In Fig. \ref{fig:4bars} the mechanism and the corresponding block diagram used for constructing the final pH system are presented. The lumped masses are directly included in the coupler and follower model considering a simple modification of the rigid mass matrix
\begin{equation}
\mathbf{M}_{rr}^{i + m_l}[1:2,1:2] = \mathbf{M}_{rr}^{i}[1:2,1:2] + \mathbf{I}_{2\times 2} m_l,
\end{equation} 
where $i=2,3$ denotes the coupler or follower model. Given a certain crank angle $\theta_1$ the relative angles between the different links are found by solving the two kinematic constraints
\begin{align*}
L_1 \cos(\theta_1)+ L_2 \cos(\theta_1+\theta_2)+ L_3 \cos(\theta_1+\theta_2+\theta_3) &=L_0, \\
L_1 \sin(\theta_1)+L_2 \sin(\theta_1+\theta_2)+L_3 \sin(\theta_1+\theta_2+\theta_3) &=0.
\end{align*} 
Once the angles describing the geometrical configuration are known, the transformer interconnection \eqref{eq:int_hinge} is applied to insert a revolute joint between adjacent links. For the deformation field a cantilever condition is imposed for each beam. The resulting system is then constrained to ground by imposing to following equalities
\begin{equation*}
\mathbf{v}_{P_1} = 0, \quad \omega^z_{P_1} = 0, \quad \mathbf{v}_{C_3} = 0.
\end{equation*}
The resulting system is expressed in pH form as $\mathbf{E}\dot{\mathbf{e}} = \mathbf{J} \mathbf{e}$. The eigenfrequencies are then found by solving the generalized eigenvalue problem $\mathbf{E}\bm{\Phi} = \mathbf{J} \bm{\Phi \Lambda}$. Since $\mathbf{J}$ is skew-symmetric the eigenvalues will be imaginary $\bm{\Lambda} = j \bm{\Omega}$. The first three pulsations  are reported in Fig. \ref{fig:omega_4bars} for different values of the crank angle $\theta_1$. The results match perfectly \cite{KITIS1990267,Chebbi2017}, assessing the validity of the linear model.

\subsection{Rotating crank-slider}
To verify the non-linear planar model a crank-slider rotating at high speed is considered. The example is retrieved form \cite{Ellenbroek2018}.  The crank is considered as rigid, with length $L_{\text{cr}} = 0.15 \ [\mathrm{m}]$ and rotates at a constant angular rate $\omega_{\text{cr}} = 150 \ [\mathrm{rad/s}]$. The flexible coupler has length $L_{\text{cl}} = 0.3 \ [\mathrm{m}]$ and a circular cross section whose diameter is $d_{\text{cl}} = 6 \ [\mathrm{mm}]$. Its Young modulus and density are given by $E_{\text{cl}}=0.2 \ 10^{12} \ [\mathrm{Pa}]$, and $\rho_{\text{cl}}=7870 \ [\mathrm{kg/m}^3]$. The slider has a total mass equal to half the mass of the coupler $m_{\text{sl}} = 0.033 \ [\mathrm{kg}]$. A simply supported condition is supposed for the coupler deformation field. This choice is motivated by the fact that the slider has a large inertia and does not allow elastic displacement at the tip.

\begin{figure}[tb]
	\centering
	\includegraphics[width=0.5\textwidth]{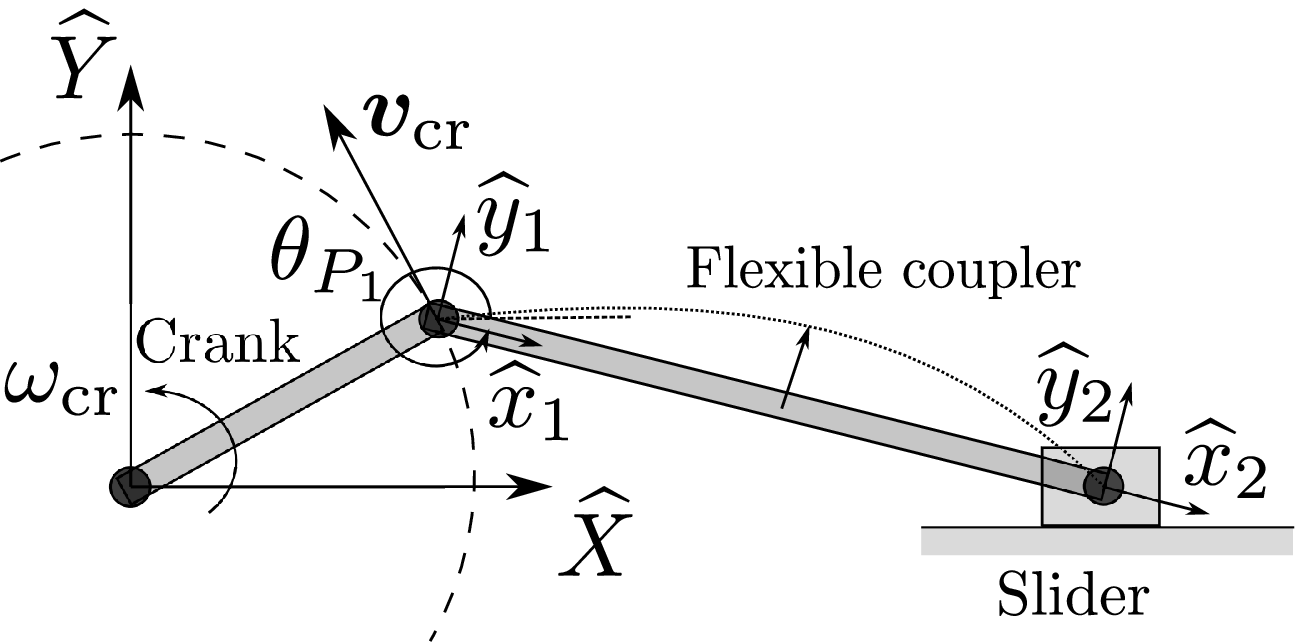} 
	\includegraphics[width=0.4\textwidth]{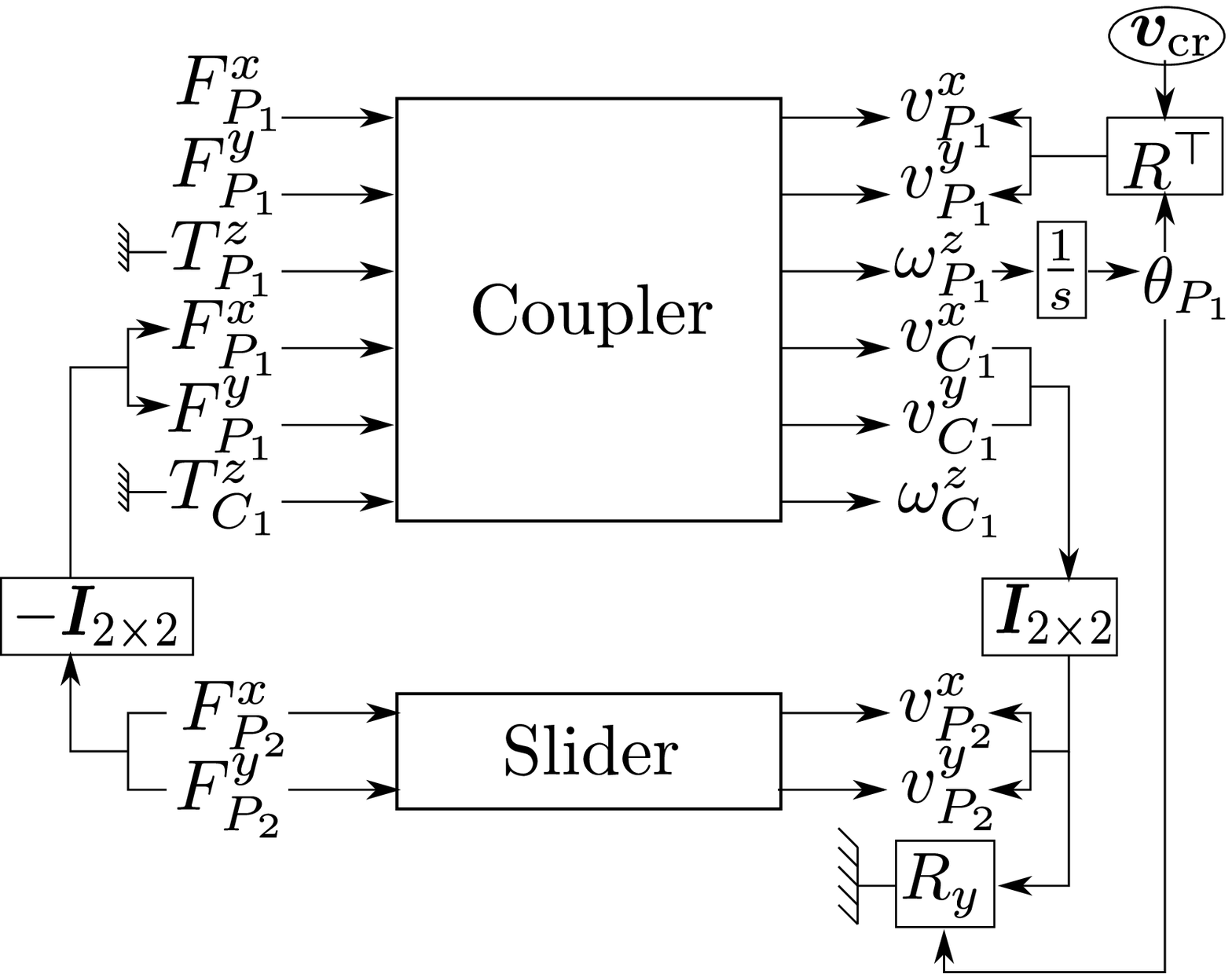} 
	\caption{Crank slider illustration (left) and block diagram (right)}
	\label{fig:crsl}
\end{figure}

An illustration of the system and the block diagram used to construct the model are provided in Fig. \ref{fig:crsl}. To construct the crank slider a transfomer interconnection is first used to connect the slider to the flexible coupler. The motion of the slider is then computed in the coupler reference frame. Then the sliding constraint, that requires the vertical velocity of the slider to be null in the inertial frame, is imposed as follows
\[
0 = \sin(\theta_{P_1}) v^x_{P_2} + \cos(\theta_{P_1}) v^y_{P_2} = \mathbf{R}_y(\theta_{P_1}) \mathbf{v}_{P_2},
\]
where $\mathbf{R}_y$ is the second line of the rotation matrix and ${\theta}_{P_1}$ is the angle defining the orientation of the coupler. The rigid crank velocity at the endpoint  
\begin{equation*}
\mathbf{v}_{\text{cr}}(t) = -\omega_{\text{cr}} L _{\text{cr}} \sin(\omega_{\text{cr}} t) \widehat{\mathbf{X}} + \omega_{\text{cr}} L _{\text{cr}} \cos(\omega_{\text{cr}} t) \widehat{\mathbf{Y}}
\end{equation*} 
has to be written in the coupler reference frame to get the input
\begin{equation}
\mathbf{u}_{\text{cl}} = R(\theta_{P_1})^\top \mathbf{v}_{\text{cr}}.
\end{equation}
The resulting system is a quasi linear index-2 DAE of the form
\begin{equation*}
\begin{bmatrix}
\mathbf{M} & 0 & 0 \\
0 & 0 & 0 \\
0 & 0 & 0 \\
\end{bmatrix}
\begin{bmatrix}
\dot{\mathbf{e}} \\ \dot{\bm{\lambda}}_0 \\ \dot{\bm{\lambda}}_u \\
\end{bmatrix}= 
\begin{bmatrix}
\mathbf{J}(\mathbf{e}) & \mathbf{G}_0^\top(\theta_{P_1}) & \mathbf{G}_u^\top \\
-\mathbf{G}_0(\theta_{P_1}) & 0 & 0 \\
-\mathbf{G}_u & 0 & 0 \\
\end{bmatrix}
\begin{bmatrix}
\mathbf{e} \\ \bm{\lambda}_0 \\ \bm{\lambda}_u \\
\end{bmatrix} + 
\begin{bmatrix}
0 \\ 0 \\ R(\theta_{P_1})^\top \\
\end{bmatrix}
\mathbf{v}_{\text{cr}}.
\end{equation*}

\begin{figure}[tb]
	\centering
	\includegraphics[width=0.45\textwidth]{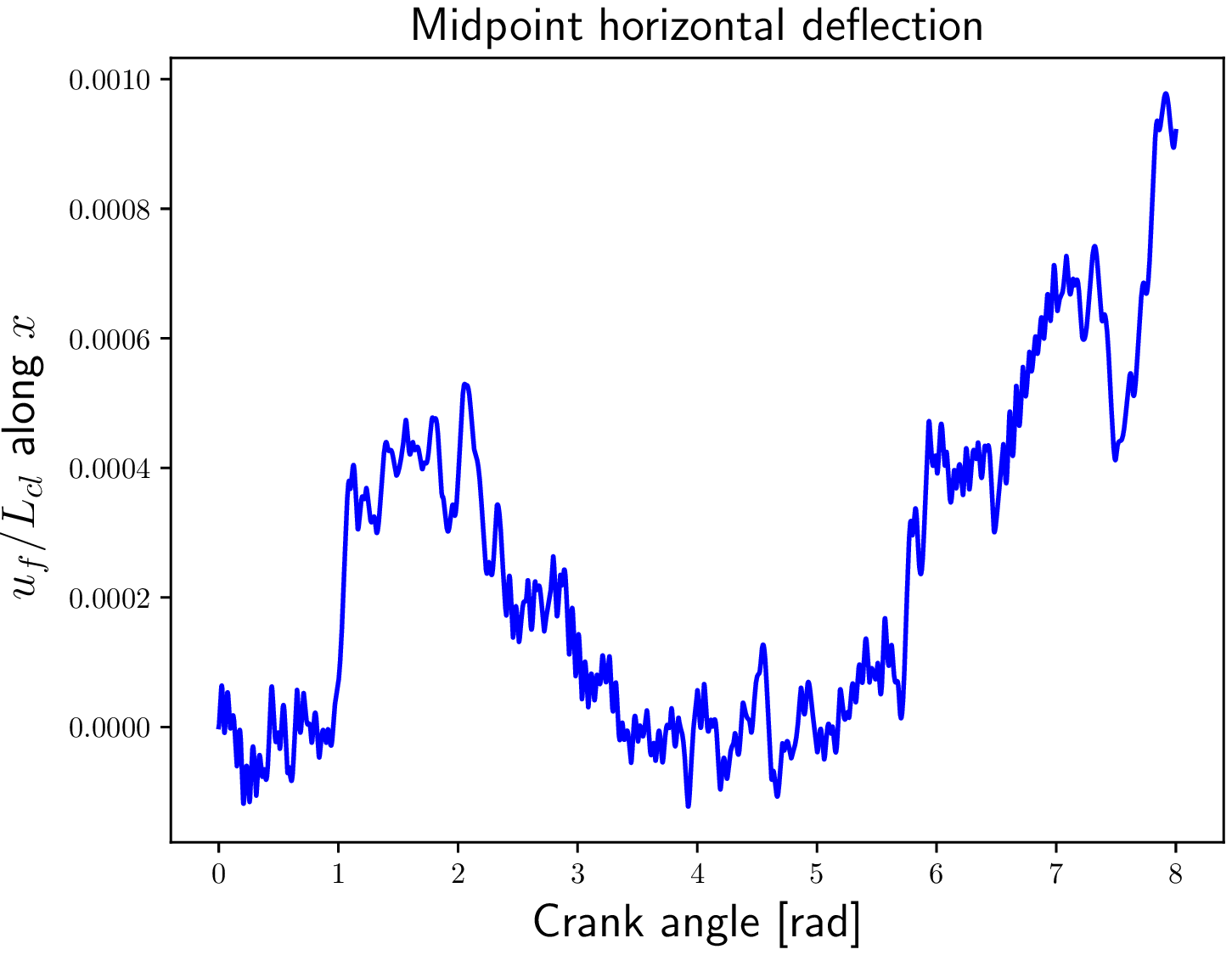} 
	\includegraphics[width=0.45\textwidth]{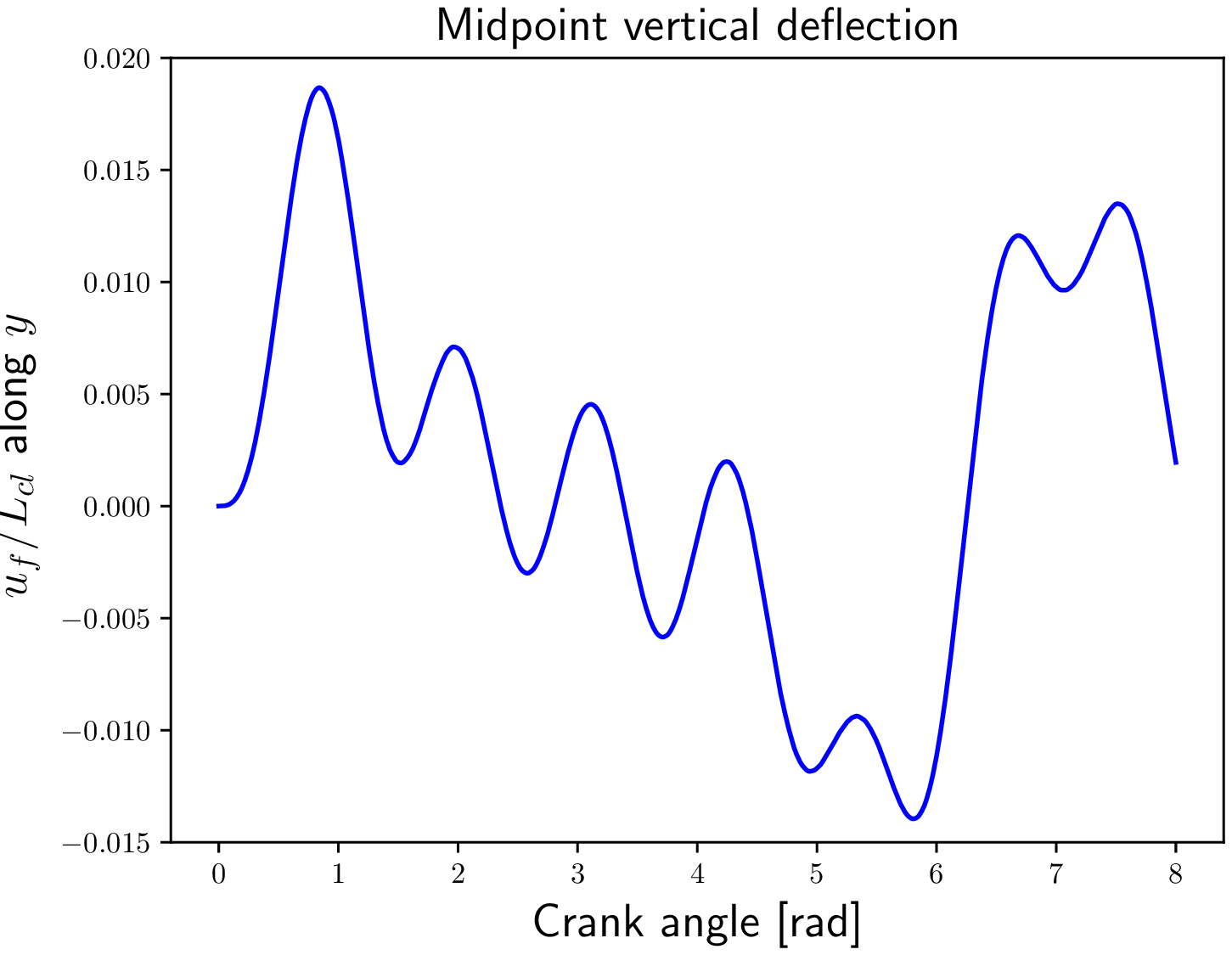} 
	\caption{Coupler midpoint horizontal (left) and vertical (right) displacement}
	\label{fig:defM_crsl}
\end{figure}

Setting the initial conditions properly is of utmost  importance for a DAE solver. For this problem the beam is supposed undeformed at the initial time. The initial conditions for the rigid movement are then found using basic kinematics considerations.  The system is then solved using the IDA algorithm available in the Assimulo library \cite{assimulo}. In Fig. \ref{fig:defM_crsl} the midpoint deformation displacement $u_f^x(L_{\text{cl}/2}),\; u_f^y(L_{\text{cl}/2})$, normalized with respect to the coupler length, is reported. The resulting vertical displacement is in accordance with the results presented in \cite{Ellenbroek2018}. The horizontal displacement exhibits high oscillations because of the higher eigenfrequencies of the longitudinal movement. This is due to the fact that null initial conditions are imposed on the deformation \cite{MB_Daepde}. In order to obtain a smoother solution, the initial deformation has to be computed from the rigid initial condition.

\subsection{Hinged spatial beam}
\begin{figure}[tb]
	\centering
	\includegraphics[width=0.45\textwidth]{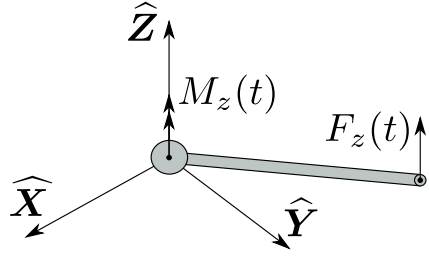} 
	\caption{Spatial beam on a spherical joint.}
	\label{fig:beam_3D}
\end{figure}

\begin{figure}[tb]
	\centering
	\includegraphics[width=0.45\textwidth]{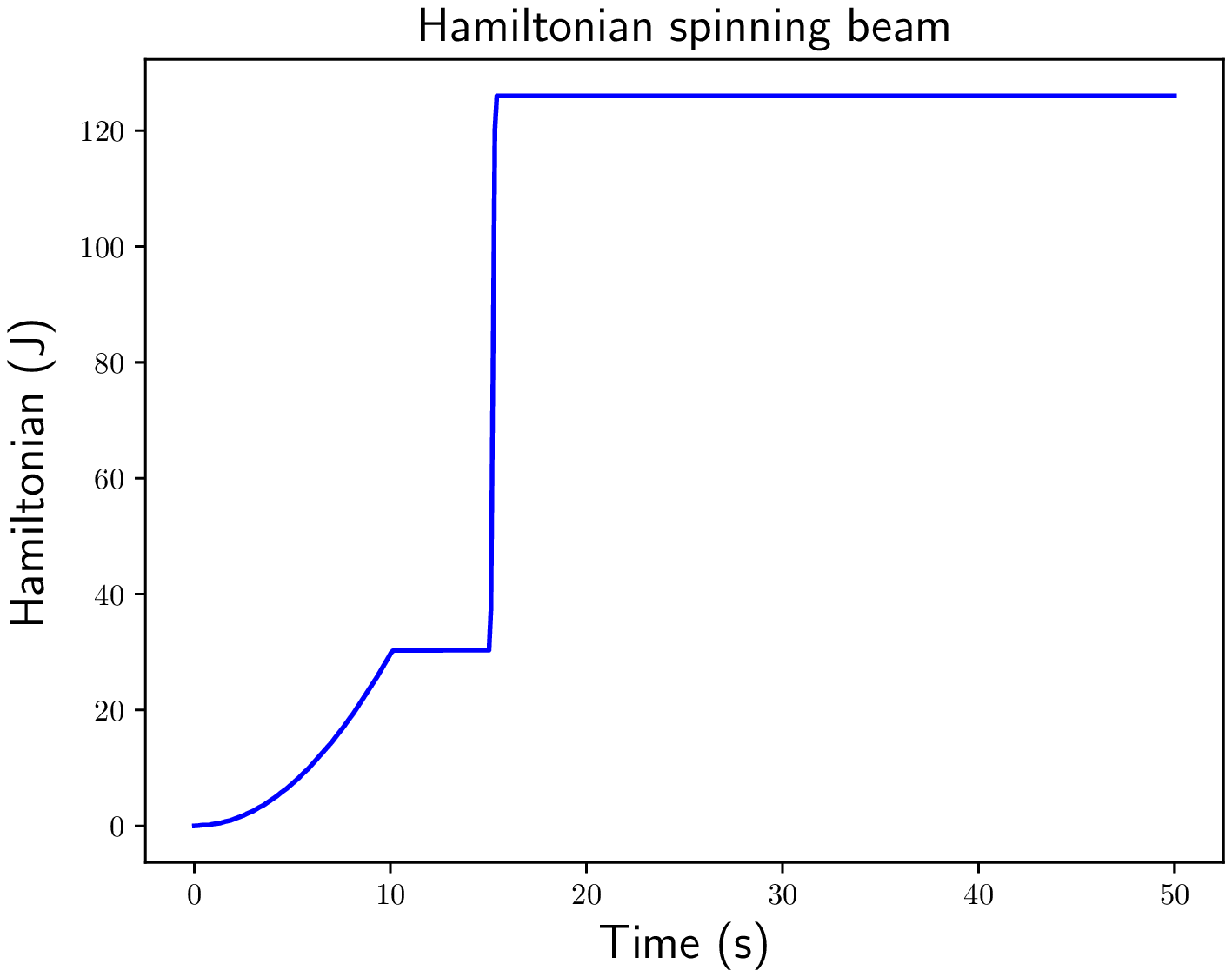} 
	\includegraphics[width=0.45\textwidth]{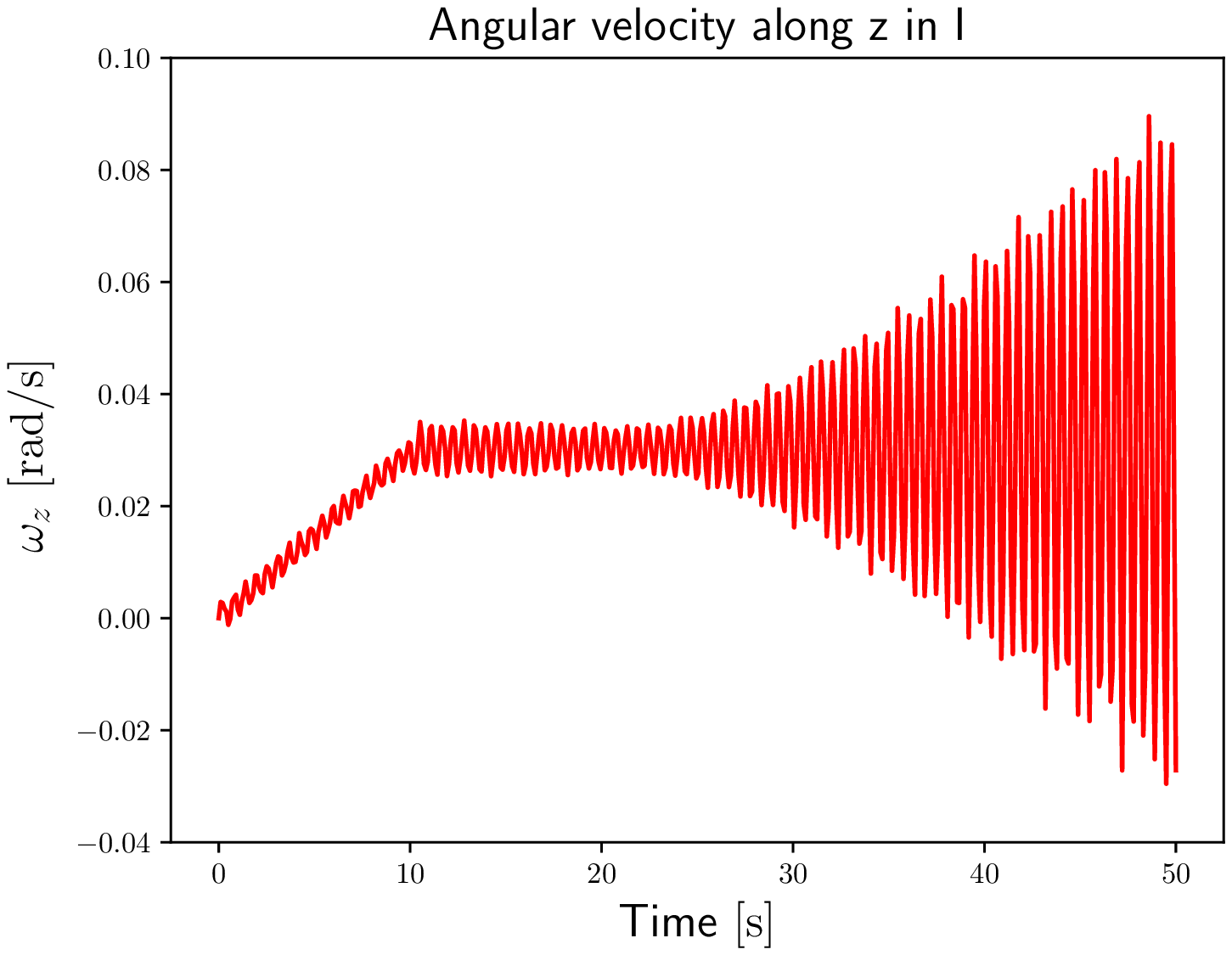} 
	\caption{Simulation results: kinetic energy (left) and angular velocity about the vertical inertial direction (right).}
	\label{fig:H_omega}
\end{figure}

A spatial beam rotating about a spherical joint is considered (see Fig. \ref{fig:beam_3D}). This example was considered in \cite{Cardona2000,Ellenbroek2018}. The physical parameters are briefly recalled in Table \ref{tab:data_3Dbeam}. The spherical joint constraint is imposed by setting to zero the linear velocity, while a cantilever is imposed for the deformation field as the tip is free. For the first $10.2 [\mathrm{s}]$ a torque $ M_z =200 [\mathrm{N/mm}]$ is applied about the vertical axis. Then, an impulsive force $ F_z =100 [\mathrm{N}]$ is applied at the tip of the beam at $15 [\mathrm{s}]$, to excite the out-of-plane movement. The system is solved using an implicit Runge-Kutta method of the Radau IIA family. The simulation results, provided in Fig. \ref{fig:H_omega}, correspond to the total energy and the angular velocity measured in the inertial vertical direction. The result matches with the provided references. Indeed the non-linearities associated to the gyroscopic terms are small as the maximum angular velocity is equal to $0.1 \ [\mathrm{rad/s}] \approx 5 \ [\mathrm{deg/s}]$. 

\begin{table}[tb]
	\caption{Physical parameters for the hinged spatial beam.}
	\label{tab:data_3Dbeam}       
	\begin{tabular}{lllll}
		\hline\noalign{\smallskip}
		Length & Cross section & Inertia moment & Density & Young modulus \\
		\noalign{\smallskip}\hline\noalign{\smallskip}
		141.45 $[\mathrm{mm}]$ & 9.0 $[\mathrm{mm}^2]$ & 6.75 $[\mathrm{mm}^4]$ & 7800 $[\mathrm{kg/mm}^3]$ & 2.1 $10^6 \ [\mathrm{N/m}^2]$  \\
		\hline
	\end{tabular}
\end{table}

\section{Conclusion}
A port-Hamiltonian formulation for the flexible multibody dynamics has been discussed. The proposed methodology, being based on a floating frame formulation, relies on the hypothesis of small deformations. However, the geometric stiffening effect can be accounted for by considering a corresponding energy. The discretization procedure uses a mixed finite element method, hence, the stress distribution is available without any post-processing. This is a valuable characteristic of this framework, as the stress distribution is the most important variable for preliminary analysis of mechanical components. Moreover, this approach allows treating different models (e.g. plates, shells) easily and in a common framework. The construction of complex multibody system becomes completely modular and well suited for control applications.  \\
Many future directions are to be investigated. Large deformation could be included by employing a substructuring technique \cite{SHABANA_substructure}.
The stability and numerical convergence of the associated finite element is still to be proved. Another interesting topic is the application of model reduction techniques. While for linear pHDAE systems consolidated methodologies exist, for the general non linear differential-algebraic case, solutions are not yet available. Numerical methods capable of preserving important structural properties in discrete time have been studied for rigid body dynamics \cite{celledoni2018passivity} and generic ODE \cite{KOTYCZKA_dt} and DAE \cite{mehrmann2019structurepreserving} pH systems. The effectiveness of those with respect to the proposed formulation has to be demonstrated.  The inclusion of control strategies is an important topic to be explored in the future.


\section*{Appendix A: Mathematical tools}
We recall here some identities and definitions that will be used throughout the paper. 
\subsection*{\normalsize \textbf{A.1 Properties of the cross product}}
We denote by $\crmat{\bm{a}}$ the skew symmetric map associated to vector $\bm{a} = [a_x, a_y, a_z]^\top$
\begin{equation}
\crmat{\bm{a}} = 
\begin{bmatrix}
0 & -a_z & a_y \\
a_z & 0 & -a_x \\
-a_y & a_x & 0 \\
\end{bmatrix}
\end{equation}
This map allows rewriting the cross product as a matrix vector product $\bm{a}\wedge \bm{b} = \crmat{\bm{a}}\bm{b}$. The cross product satisfies the anticommutativity property
\begin{equation}
\label{eq:anticom}
\crmat{\bm{a}} \bm{b} = - \crmat{\bm{b}} \bm{a}, \qquad \bm{a}, \bm{b} \in \mathbb{R}^3.
\end{equation}
Furthermore, it satisfies the Jacobi Identity
\begin{equation}
\label{eq:jacobi}
\crmat{\bm{a}} (\crmat{\bm{b}} \bm{c}) + \crmat{\bm{b}} (\crmat{\bm{c}} \bm{a}) + \crmat{\bm{c}} (\crmat{\bm{a}} \bm{b}) = 0, \qquad \bm{a}, \bm{b}, \bm{c} \in \mathbb{R}^3.
\end{equation}

\subsection*{\normalsize \textbf{A.2 Adjoint of operators}}
In this paper, the adjoint of an operator is used. We recall the necessary definitions.
\begin{definition}
	Given a linear operator $\mathcal{A}: \mathscr{H}^1 \rightarrow \mathscr{H}^2$ between Hilbert spaces, the adjoint $\mathcal{A}^*:\mathscr{H}^2\rightarrow~\mathscr{H}^1$ fulfills 
	\begin{equation}
	\langle y, \mathcal{A}x \rangle_{\mathscr{H}^2} = \langle \mathcal{A}^* y, x \rangle_{\mathscr{H}^1}, \qquad x \in \mathscr{H}^1, y \in \mathscr{H}^2.
	\end{equation}
\end{definition}
To illustrate this definition, consider the operator $\mathcal{I}^\Omega = \int_\Omega (\cdot) \d\Omega : \mathscr{L}^2(\Omega, \mathbb{R}^3) \rightarrow \mathbb{R}^3$. Given a function $\bm{u} \in \mathscr{L}^2(\Omega, \mathbb{R}^3)$ and a vector $\bm{v} \in \mathbb{R}^3$, then the adjoint operator $(\mathcal{I}^\Omega)^*$ extends the vector $\bm{v}$ as a constant vector field over $\Omega$
\begin{equation*} 
\langle \bm{v}, \mathcal{I}^\Omega \bm{u} \rangle_{\mathbb{R}^3} = \langle (\mathcal{I}^\Omega)^* \bm{v},  \bm{u} \rangle_{\mathscr{L}^2(\Omega, \mathbb{R}^3)}. 
\end{equation*}
\begin{definition}
	A linear bounded operator $\mathcal{A}: \mathscr{H} \rightarrow \mathscr{H}$ is self-adjoint if it holds
	\begin{equation}
	\langle y, \mathcal{A}x \rangle_{\mathscr{H}} = \langle \mathcal{A} y, x \rangle_{\mathscr{H}}, \qquad x, y \in \mathscr{H}.
	\end{equation}
\end{definition}
\begin{definition}
	A linear bounded operator $\mathcal{A}: \mathscr{H} \rightarrow \mathscr{H}$ is  skew-adjoint if it holds
	\begin{equation}
	\langle y, \mathcal{A}x \rangle_{\mathscr{H}} = -\langle \mathcal{A} y, x \rangle_{\mathscr{H}}, \qquad x, y \in \mathscr{H}.
	\end{equation}	
\end{definition}
Indeed, the differential operators that appears in $\bm{\mathcal{J}}$ ($\Div, \Grad$), are unbounded in the $\mathscr{L}^2$ topology. Whenever unbounded operators are considered, it is important to define their domain. To avoid the need of specifying domains, the notion of formal (or essential) adjoint can be evoked. The formal adjoint respects the integration by parts formula and is defined only for sufficiently smooth functions with compact support. In this sense $\Div, \Grad$ are formally skew-adjoint, since for smooth functions with compact support, it holds
\begin{equation*}
\left\langle y, \, \Grad(x) \right\rangle_{\mathscr{L}^2(\Omega, \mathbb{R}^{3\times 3}_{\text{sym}})} 
\underbrace{=}_{\text{I.B.P.}} -\left\langle\Div(y), \, x \right\rangle_{\mathscr{L}^2(\Omega, \mathbb{R}^3)}.
\end{equation*}
The definition of the domain of the operators, that requires the knowledge of the boundary conditions, has not been specified. For this reason, the $\bm{\mathcal{J}}$ operator is said to be formally skew-adjoint (or simply skew-symmetric).

\subsection*{\normalsize \textbf{A.3 Index of a differential-algebraic system}}
When dealing with differential-algebraic systems an important notion is the index.
\begin{definition}
	The index of a DAE is the minimum number of differentiation steps required to transform a DAE into an ODE.
\end{definition}
Because of their structure, pH multibody systems are of index two. Consider for simplicity a generic linear pH multibody system, whose equations are
\begin{equation*}
\begin{aligned}
\mathbf{M} \dot{\mathbf{e}} &=  \mathbf{J}\mathbf{e} + \mathbf{G}^\top \bm{\lambda} + \mathbf{B}\mathbf{u}, \\ 
0 &= -\mathbf{G}\mathbf{e}.
\end{aligned}
\end{equation*}
Matrix $\mathbf{M}$ is squared and invertible and matrix $\mathbf{G}$ is full rank. If the second equation is derived twice in time, then it is obtained
\[\dot{\bm{\lambda}} = - (\mathbf{G} \mathbf{M}^{-1} \mathbf{G}^\top)^{-1} \mathbf{G} \mathbf{M}^{-1} (\mathbf{J} \dot{\mathbf{e}} + \mathbf{B}\dot{\mathbf{u}}).
\]
Therefore, the system index is two.  
\section*{Appendix B: Detailed derivation of the equation of motions}

The detailed derivation of the pH system \eqref{eq:ph_mfd_all} is presented here. We stick to the notation adopted along the paper. First, let us recall the equations for a floating flexible body reported in \cite{MB_Daepde,simeon2013computational}.
\begin{itemize}
	\item Linear momentum balance:
	\begin{equation}
	\label{eq:rig_tr3}
	\begin{split}
	&m ^{i}\ddot{\bm{r}}_P + \bm{R} \crmat{\bm{s}_u}^\top \dot{\bm{\omega}}_P  + \bm{R} \int_{\Omega} \rho \ddot{\bm{u}}_f \d{\Omega} = \\
	&\quad + \bm{R} \left\{ -\crmat{\bm{\omega}_P} \crmat{\bm{\omega}_P} \bm{s}_u -  \int_{\Omega} 2 \rho \crmat{\bm{\omega}_P} \dot{\bm{u}}_f \d{\Omega} +  \int_{\Omega} \bm\beta \d{\Omega} +  \int_{\partial \Omega} \bm\tau \d{\Gamma}  \right\} 
	\end{split}
	\end{equation}
	\item Angular momentum balance:
	\begin{equation}
	\label{eq:rig_rot3}
	\begin{split}
	\crmat{\bm{s}_u} {\bm{R}^{\top}} \ ^{i}\ddot{\bm{r}}_P + \bm{J}_u \dot{\bm\omega}_P + \int_{\Omega} \rho \crmat{\bm{x}+\bm{u}_f} \ddot{\bm{u}}_f \d{\Omega} + \crmat{\bm{\omega}_P} \bm{J}_u \bm{\omega}_P = \\ 
	- \int_{\Omega} 2\rho \crmat{\bm{x}+\bm{u}_f} \crmat{\bm\omega_P} \dot{\bm{u}}_f \d{\Omega} + \int_{\Omega}\crmat{\bm{x}+\bm{u}_f} \bm\beta \d{\Omega} + \int_{\partial \Omega}\crmat{\bm{x}+\bm{u}_f} \bm\tau \d{\Gamma} \\
	\end{split}
	\end{equation}
	\item Flexibility PDE:
	\begin{equation}
	\label{eq:flex3}
	\rho  {\bm{R}^{\top}} \ ^{i}\ddot{\bm{r}}_P + \rho (\crmat{\dot{\bm\omega}_P} + \crmat{\bm{\omega}_P}\crmat{\bm{\omega}_P})(\bm{x}+\bm{u}_f) + \rho (2 \crmat{\bm{\omega}_P} \dot{\bm{u}}_f + \ddot{\bm{u}}_f) = \Div{\bm\Sigma} + \bm\beta,
	\end{equation}
\end{itemize}
The first two equations are written in the inertial frame and so they need to be projected in the body frame. Considering that the position of point $P$, i.e. $^{i}{\bm{r}}_P$, is computed in the inertial frame and $\bm{v}_P$ in the body frame, it holds $^{i}\dot{\bm{r}}_P = \bm{R} \bm{v}_P$. The derivative of this gives
\begin{equation}
\label{eq:acc_rP}
^{i}\ddot{\bm{r}}_P = \bm{R} \left(\dot{\bm{v}}_P + \crmat{\bm{\omega}_P} \bm{v}_P \right)
\end{equation}
If \eqref{eq:acc_rP} is put into \eqref{eq:rig_tr3}, \eqref{eq:rig_rot3}, \eqref{eq:flex3} and pre-multiplying  Eq. \eqref{eq:rig_tr3} by $\bm{R}^\top$, Eqs. \eqref{eq:rig_tr1} \eqref{eq:rig_rot1}, \eqref{eq:flex1} are obtained.
\begin{itemize}
	\item Linear momentum balance:
	\begin{equation}
	\label{eq:rig_tr4}
	\begin{split}
	&m (\dot{\bm{v}}_P + \crmat{\bm{\omega}_P} \bm{v}_P) + \crmat{\bm{s}_u}^\top \dot{\bm{\omega}}_P  + \int_{\Omega} \rho \dot{\bm{v}}_f \d{\Omega} = \\
	&\quad - \crmat{\bm{\omega}_P} \crmat{\bm{\omega}_P} \bm{s}_u - \int_{\Omega} 2 \rho \crmat{\bm{\omega}_P} {\bm{v}}_f \d{\Omega} +  \int_{\Omega} \bm\beta \d{\Omega} + \int_{\partial \Omega} \bm\tau \d{\Gamma}.
	\end{split}
	\end{equation}
	\item Angular momentum balance:
	\begin{equation}
	\label{eq:rig_rot4}
	\begin{split}
	\crmat{\bm{s}_u} (\dot{\bm{v}}_P + \crmat{\bm{\omega}_P} \bm{v}_P) + \bm{J}_u \dot{\bm\omega}_P + \int_{\Omega} \rho \crmat{\bm{x}+\bm{u}_f} \dot{\bm{v}}_f \d{\Omega} + \crmat{\bm{\omega}_P} \bm{J}_u \bm{\omega}_P = \\ 
	- \int_{\Omega} 2\rho \crmat{\bm{x}+\bm{u}_f} \crmat{\bm\omega_P} {\bm{v}}_f \d{\Omega} + \int_{\Omega}\crmat{\bm{x}+\bm{u}_f} \bm\beta \d{\Omega} + \int_{\partial \Omega}\crmat{\bm{x}+\bm{u}_f} \bm\tau \d{\Gamma}. \\
	\end{split}
	\end{equation}
	\item Flexibility PDE:
	\begin{equation}
	\label{eq:flex4}
	\begin{split}
	\rho (\dot{\bm{v}}_P + \crmat{\bm\omega_P} \bm{v}_P) + \rho (\crmat{\dot{\bm\omega}_P} + \crmat{\bm{\omega}_P}\crmat{\bm{\omega}_P})(\bm{x}+\bm{u}_f) + \rho (2 \crmat{\bm{\omega}_P} {\bm{v}}_f + \dot{\bm{v}}_f) = \\
	\Div{\bm\Sigma} + \bm\beta,
	\end{split}
	\end{equation}
	where $\bm{v}_f = \dot{\bm{u}}_f$.
\end{itemize}

Consider now the term $\crmat{\bm{\omega}_P} (\crmat{\bm{\omega}_P} \bm{s}_u)$, appearing in \eqref{eq:rig_tr4}. Using the anticommutativity \eqref{eq:anticom} and the fact that the cross map is skew-symmetric $\crmat{\bm{a}} = -\crmat{\bm{a}}^\top$ one finds
\begin{align*}
-\crmat{\bm{\omega}_P} (\crmat{\bm{\omega}_P} \bm{s}_u) = \crmat{\crmat{\bm{s}_u}^\top\bm{\omega}_{P}} \bm{\omega}_{P}.
\end{align*}
Eq. \eqref{eq:rig_tr4} is then rewritten as
\begin{equation}
\label{eq:rig_tr5}
\begin{split}
m\dot{\bm{v}}_P + \crmat{\bm{s}_u}^\top \dot{\bm{\omega}}_P +   \int_{\Omega} \rho \dot{\bm{v}}_f \d{\Omega}  = \\
\left[m \bm{v}_P + \crmat{\bm{s}_u}^\top \bm\omega_P +2 \int_{\Omega} \rho \bm{v}_f \d{\Omega} \right]_\times \bm\omega_P +  \int_{\Omega} \bm\beta \d{\Omega} + \int_{\partial \Omega} \bm\tau \d{\Gamma}.
\end{split}
\end{equation}
The terms $\crmat{\bm{s}_u} (\crmat{\bm{\omega}_P} \bm{v}_P), \; 2\rho \crmat{\bm{x}+\bm{u}_f} (\crmat{\bm\omega_P} {\bm{v}}_f)$, appearing in \eqref{eq:rig_rot4} can be rewritten using the Jacobi identity \eqref{eq:jacobi}
\begin{align}
\crmat{\bm{s}_u} (\crmat{\bm{\omega}_P} \bm{v}_P) &= - \crmat{\crmat{\bm{s}_u} \bm{v}_P} \bm{\omega}_P - \crmat{\crmat{\bm{s}_u}^\top \bm{\omega}_P} \bm{v}_P, \\
2\rho \crmat{\bm{x}+\bm{u}_f} (\crmat{\bm\omega_P} {\bm{v}}_f) &= - \crmat{2\rho \crmat{\bm{x}+\bm{u}_f} \bm{v}_f} \bm\omega_P - \crmat{2\rho \crmat{\bm{x}+\bm{u}_f}^\top \bm\omega_P} \bm{v}_f
\end{align}
Eq. \eqref{eq:rig_rot4} is then rewritten as
\begin{equation}
\label{eq:rig_rot5}
\begin{split}
\crmat{\bm{s}_u} \dot{\bm{v}}_P  + \bm{J}_u \dot{\bm\omega}_P + \int_{\Omega} \rho \crmat{\bm{x}+\bm{u}_f} \dot{\bm{v}}_f \d{\Omega} = \\
\left[\crmat{\bm{s}_u}^\top \bm\omega_P + 2 \int_{\Omega} \rho \bm{v}_f \d{\Omega} \right]_\times \bm{v}_P + \left[\crmat{\bm{s}_u} \bm{v}_P + \bm{J}_u \bm\omega_P + 2 \int_{\Omega} \rho \crmat{\bm{x}+\bm{u}_f} {\bm{v}}_f \d{\Omega} \right]_\times \bm\omega_P + 
\\
2 \int_{\Omega} \left[\rho \bm{v}_P + \rho \crmat{\bm{x}+\bm{u}_f}^\top \, \bm\omega_P \right]_\times \bm{v}_f \d{\Omega} + \int_{\Omega}\crmat{\bm{x}+\bm{u}_f} \bm\beta \d{\Omega} + \int_{\partial \Omega}\crmat{\bm{x}+\bm{u}_f} \bm{\tau} \d{\Gamma}.
\end{split}
\end{equation}
Notice that $2 \crmat{\bm{v}_f}\bm{v}_P + 2 \crmat{\bm{v}_P}\bm{v}_f = 0$. Using again the anticommutativity Eq. \eqref{eq:flex4} is expressed as 
\begin{equation}
\label{eq:flex5}
\begin{split}
\rho \dot{\bm{v}}_P + \rho \crmat{\bm{x}+\bm{u}_f}^\top \dot{\bm\omega}_P  + \rho \dot{\bm{v}}_f = \\
\left[\rho \bm{v}_P + \rho \crmat{\bm{x}+\bm{u}_f}^\top \bm\omega_P + 2 \rho \bm{v}_f \right]_\times \bm\omega_P + \Div{\bm\Sigma} + \bm\beta.
\end{split}
\end{equation}
Indeed, Eqs. \eqref{eq:rig_tr5}, \eqref{eq:rig_rot5}, \eqref{eq:flex5} are exactly \eqref{eq:rig_tr2}, \eqref{eq:rig_rot2}, \eqref{eq:flex2}. Now by definitions \eqref{eq:mod_momenta}, \eqref{eq:mod_momentafl}
\begin{align*}
\widetilde{\bm{p}}_t &= m \bm{v}_P + \crmat{\bm{s}_u}^\top \bm\omega_P +2 \int_{\Omega} \rho \bm{v}_f \d{\Omega}, \\
\widetilde{\bm{p}}_r &= \crmat{\bm{s}_u} \bm{v}_P + \bm{J}_u \bm\omega_P + 2 \int_{\Omega} \rho \crmat{\bm{x}+\bm{u}_f} {\bm{v}}_f \d{\Omega}, \\
\bm{\mathcal{I}}_{p_f}^\Omega(\cdot) &= \int_{\Omega} \left[ 2 \left(\rho \bm{v}_P + \rho \crmat{\bm{x}+\bm{u}_f}^\top \, \bm\omega_P +\rho \bm{v}_f\right) + \rho \bm{v}_f \right]_\times (\cdot) \d{\Omega}, 
\end{align*}
Eqs. \eqref{eq:rig_tr3}, \eqref{eq:rig_rot3}, \eqref{eq:flex3} are written as 
\begin{equation}
\bm{\mathcal{M}}
\diff{}{t}
\begin{bmatrix}
\bm{v}_P \\ \bm\omega_P  \\ \bm{v}_f  \\ \bm\Sigma \\
\end{bmatrix} = 
\begin{bmatrix}
0 & \crmat{\widetilde{\bm{p}}_t} & 0 & 0 \\
\crmat{\widetilde{\bm{p}}_t} & \crmat{\widetilde{\bm{p}}_r} & \bm{\mathcal{I}}_{p_f}^\Omega & 0 \\
0 & -(\bm{\mathcal{I}}_{p_f}^\Omega)^* & 0 & \Div \\
0 & 0 & \Grad & 0 \\
\end{bmatrix}
\begin{bmatrix}
\bm{v}_P \\ \bm\omega_P  \\ \bm{v}_f  \\ \bm\Sigma \\
\end{bmatrix} - 
\begin{bmatrix}
0 \\ 0  \\ \delta_{\bm{u}_f}H  \\ 0 \\
\end{bmatrix},
\end{equation} 
with
\begin{align*}
\bm{\mathcal{M}} &= 
\begin{bmatrix}
m \bm{I}_{3\times 3} & \crmat{\bm{s}_u}^\top & \mathcal{I}_\rho^{\Omega} & 0 \\
\crmat{\bm{s}_u} & \bm{J}_u & \bm{\mathcal{I}}_{\rho x}^{\Omega} & 0  \\
(\mathcal{I}_\rho^{\Omega})^* & (\bm{\mathcal{I}}_{\rho x}^{\Omega})^* & \rho & 0  \\
0 & 0 & 0 & \bm{\mathcal{D}}^{-1} \\
\end{bmatrix}, \qquad \text{see \eqref{eq:mass_op}} \\
H &= \frac{1}{2} \int_{\Omega} \left\{\rho ||\bm{v}_P + \crmat{\bm{\omega}_P} (\bm{x}+\bm{u}_f) + {\bm{v}}_f||^2 + \bm\Sigma \cddot \bm\varepsilon \right\}  \d{\Omega}, \qquad \text{see \eqref{eq:H}}.
\end{align*}
Hence, it is clear that Eqs. \eqref{eq:rig_tr3}, \eqref{eq:rig_rot3}, \eqref{eq:flex3} from \cite{MB_Daepde,simeon2013computational} are equivalently recast in the form \eqref{eq:ph_mfd_all}.
\bibliographystyle{spmpsci}      
\bibliography{multibody} 


\end{document}